\documentstyle[12pt]{article}
\begin{document}
\renewcommand{\theequation}{\thesection.\arabic{equation}}
\input epsf.tex

%  Greek letters
\def\a{\alpha}
\def\b{\beta}
\def\ch{\chi}
\def\d{\delta}
\def\e{\epsilon}
\def\E{{\cal E}}
\def\f{\phi}
\def\g{\gamma}
\def\h{\eta}
\def\i{\iota}
\def\j{\psi}
\def\k{\kappa}
\def\l{\lambda}
\def\m{\mu}
\def\n{\nu}
\def\o{\omega}
\def\p{\pi}
\def\q{\theta}
\def\r{\rho}
\def\s{\sigma}
\def\t{\tau}
\def\u{\upsilon}
\def\x{\xi}
\def\z{\zeta}
\def\D{\Delta}
\def\F{\Phi}
\def\G{\Gamma}
\def\J{\Psi}
\def\L{\Lambda}
\def\O{\Omega}
\def\P{\Pi}
\def\S{\Sigma}
\def\U{\Upsilon}
\def\X{\Xi} 
\def\T{\Theta}
\def\vf{\varphi}
\def\ve{\varepsilon}
\def\cC{{\cal X}}
\def\cD{{\cal Y}}

\def\Ab{\bar{A}}
\def\gi{g^{-1}}
\def\li{{ 1 \over \l } }
\def\lb{\l^{*}}
\def\zb{\bar{z}}
\def\ub{u^{*}}
\def\vb{v^{*}}
\def\Tb{\bar{T}}
\def\pp {\partial }
\def\pb {\bar{\partial }}
\def\be{\begin{equation}}
\def\ee{\end{equation}}
\def\ben{\begin{eqnarray}}
\def\een{\end{eqnarray}}
\def\lt{\tilde{\lambda}}
\hsize=15truecm
\addtolength{\topmargin}{-0.6in}
\addtolength{\textheight}{1.5in}
\vsize=26.5truecm
\hoffset=-0.3in

\thispagestyle{empty}
\begin{flushright} August \ 1997\\
SNUTP 97-110 \\
% hep-th/9709002 \\
\end{flushright}
\begin{center}
 {\large\bf  Field Theory for
  Coherent Optical Pulse Propagation }
\vglue .5in
 Q-Han Park\footnote{ E-mail address; qpark@nms.kyunghee.ac.kr }
\vglue .2in
{and}
\vglue .2in
H. J. Shin\footnote{ E-mail address; hjshin@nms.kyunghee.ac.kr }
\vglue .2in
{\it  
Department of Physics \\
and \\
Research Institute of Basic Sciences \\
Kyunghee University\\
Seoul, 130-701, Korea}
\vglue .2in
{\bf ABSTRACT}\\[.2in]
\end{center}
We introduce a new notion of ``matrix potential" to nonlinear 
optical systems. In terms of a matrix potential $g$, we present a 
gauge field theoretic formulation of the Maxwell-Bloch equation that 
provides a semiclassical description of the propagation of optical pulses 
through resonant multi-level media. We show that the Bloch part of the 
equation can solved identically through $g$ and the remaining Maxwell 
equation becomes a second order differential equation with reduced set 
of variables due to the gauge invariance of the system.  Our formulation 
clarifies the (nonabelian) symmetry structure of the Maxwell-Bloch 
equations for various multi-level media in association with symmetric 
spaces $G/H$. In particular, we associate nondegenerate two-level system 
for self-induced transparency with $G/H=SU(2)/U(1)$ and three-level 
$\L $- or $V$-systems with $G/H = SU(3)/U(2)$. 
We give a detailed analysis for the two-level case in the matrix 
potential formalism, and address various new properties of the system 
including soliton numbers, effective potential energy, gauge and discrete
symmetries, modified pulse area, conserved topological and nontopological 
charges. The nontopological charge measures the amount of self-detuning of 
each pulse. Its conservation law leads to a new type of pulse stability 
analysis which explains nicely earlier numerical results.
\vglue .1in
\newpage
\tableofcontents
\section{Introduction}
\setcounter{equation}{0}
Since the invention of the laser, much progress has been made in 
understanding nonlinear interactions of radiation with matter which 
made nonlinear optics a fast developing and independent field of 
science.  Recently, the interaction of laser lights with a 
multi-level optical medium has attracted more attention in 
the context of lasing without inversion \cite{lwi,lwiexp} and 
electromagnetically induced transparency (EIT) \cite{Boller}.
Laser light in general is expressed in terms of a 
macroscopic, classical electric field which interacts with 
microscopic, quantum mechanical matter. Unlike classical 
electrodynamics, the electric scalar potential and the magnetic 
vector potential do not appear to replace electromagnetic fields 
in nonlinear optics. Instead, the electric field itself, with 
appropriate restrictions to accomodate specific physical
problems, plays the role of a fundamental variable which renders the 
problem lacking a field theoretic Lagrangian formulation. Of course, one
could setup the problem in the most general QED Lagrangian framework with 
the conventional potential variable $A_{\m}$, but the nonlinearity of 
interactions and various approximation schemes involved make the use of 
potential $A_{\m}$ meaningless. For instance, the Maxwell-Bloch equation 
which governs the interaction between radiation and matter takes a 
nonrelativistic, semiclassical limits of QED together with slowly varying 
envelope aprroximation (SVEA) and/or rotating wave approximation (RWA). 
Variables of the Maxwell-Bloch equation are given by the envelope 
functions of electric fields, and the components of the density matrix 
or the probility amplitudes for each atomic level occupations. Thus, 
all previous works have focused on the study of the Maxwell-Bloch equation 
itself, without making any reference to the Lagrangian and potential 
variables. However, there exists one notable exception.
In the case of nondegenerate two-level atoms, McCall and Hahn \cite{McCall} 
have shown that lossess propagation of light pulses, the phenomenon of 
self-induced transparency (SIT), can be explained  in terms of a 
potential-like variable $\theta (x)$, the time area of a suitably 
chosen electric field, which obeys the area theorem. 
Under certain circumstances, the system can be described by an effective 
potential variable $\vf (x,t)$ which satisfies the well-known sine-Gordon 
equation. In this case, the 1-soliton of the sine-Gordon theory is identified 
with the $2\pi $-pulse of McCall and Hahn. The cosine potential term
becomes proportional to the microscopic atomic energy, and the stability of 
the $2 \pi$-pulse is explained through the topological charge conservation law.
Recently, the quantum sine-Gordon theory has been also applied to the 
Maxwell-Bloch equation and quantum optics with interesting results 
\cite{Leclair}. However, one serious drawback of the sine-Gordon 
approach to the Maxwell-Bloch system is its oversimplification. In the
sine-Gordon limit, frequency detuning and frequency modulation effects are
all ignored and microscopic atomic motions (inhomogeneous broadening) are
not taken into account. Also, the model is limited only to the 
nondegenerate two-level case while many recent interesting applications are 
based on the multi-level (three-level and higher) and possibly degenerate 
systems. In an earlier work \cite{park1}, we have shown that even the
nondegenerate two-level system should be described by the complex sine-Gordon 
equation. This generalizes the sine-Gordon equation by including a phase 
degree of freedom which accounts for frequency modulation effects. We have
also shown that a more general framework can be given by a $2 \times 2$ 
matrix potential $g$ and its Lagrangian formulation. This allowed us to 
incoporate frequency detuning and external magnetic fields.  
Until now, the sine-Gordon theory was the only available field theory for
the Maxwell-Bloch system and therefore all analytic works beyond the
simplest two-level case have resorted to the Maxwell-Bloch equation, 
finding soliton type solutions through the inverse scattering
method in integrable cases (for a review, see \cite{maim2} and other 
references therein). Following the pioneering work
of Lamb \cite{lamb}, Ablowitz, Kaup and Newell have extended the inverse 
scattering formalism to include inhomogeneous broadening and obtained 
exact solutions \cite{AKN}. In accordance with the area theorem, these 
solutions show that an arbitrary initial pulse with sufficient strength 
decomposes into a finite number of $2\pi $ pulses and $0\pi $  pulses, 
plus radiation which decays exponentially. Extensions to the degenerate 
as well as the multi-level cases have been also found resulting more 
complicated soliton solutions \cite{maim2,maim,bash1,Bash2}. 
\\

In this paper, we introduce a new matrix potential variable $g$ to 
nonlinear optical systems described by (integrable) Maxwell-Bloch 
equations, and present a completely different type of analysis of the 
Maxwell-Bloch equation based a field theory formulation through $g$. 
We show that the Bloch part of the equation can be solved identically 
in terms of $g$ and the remaining Maxwell part becomes a second order 
differential equation in $g$. This is compared with the linear case of 
electromagnetism where the curl-free condition is solved in terms of a 
scalar potential $\vf$ and the remaining Gauss equation changes into the 
second order differential equation in $\vf$. The field theory action for 
the second order Maxwell equation in $g$ is provided by a sigma model-type 
action which combines the so-called ``the 1+1-dimensional $G/H$-gauged 
Wess-Zumino-Novikov-Witten action" with an appropriately chosen potential 
energy term. This work which generalizes the earlier work on the two-level 
case \cite{park1} to the multi-level cases uncovers many new features of 
the problem. In particular, our formulation clarifies the hidden 
(nonabelian) group structure of the multi-level Maxwell-Bloch equation 
in association with symmetric spaces $G/H$. For instance, nondegenerate 
two-level system of self-induced transparency is associated with 
$G/H=SU(2)/U(1)$ while three-level $\L $- or $V$-systems are associated 
with $G/H = SU(3)/U(2)$. These nonabelian group structures are shown to 
arise from the probability conservation law of a density matrix and also 
from the selection rules in relevant dipole transitions. In general, the 
number of degrees of freedom for the Maxwell equation (those of electric 
field components) is smaller than that of the matrix potential $g$ 
belonging to the group $G$. We show that these residual degrees can be 
removed by imposing constraints on $g$ through ``gauging" the action so 
that the action possesses the $H$-vector gauge invariance. The gauge 
transformation, however, is shown to receive physical meaning at the atomic 
level. That is, it accounts for the effects of frequency detuning and 
external magnetic fields. We show that inhomogeneous broadening can be also 
incoporated into the matrix potential formalism. 

In order to demonstrate the power of our matrix potential approach, we 
make a detailed analysis of optical pulses. This shows that the matrix potential  
not only leads to a deeper understanding of optical pulses,  but it also provides 
new solutions, new conserved charges and symmetries. In particular, a new type of 
stability analysis is made which generalizes the area theorem to a certain extent.
Specifically, we clarify the topological nature of solitary pulses through 
the effective potential energy term and its degenerate vacua. We define the 
topological soliton number according to the group structure of the system 
and show that a solitary pulse for certain multi-level cases, e.g. the 
degenerate three-level case, carry more than one soliton numbers. 
Also, we show that $2 \pi $ pulses can be nontopological carrying a nontopological 
charge. A nontopological soliton is interpreted as a ``self-detuned" $2 \pi$ 
pulse and the nontopological charge is shown to measure the amount of 
frequency self-detuning. The conservation laws of the topological and the 
nontopological charges are shown to prove the stability of pulses.
In particular, we prove the stability of $2 \pi $ pulses against small 
fluctuations. This explains nicely the frequency pulling effect in the 
presence of frequency detuning which has been predicted earlier by a 
numerical work.

Our matrix potential formalism also allows a systematic understanding of 
various symmetry structures of the Maxwell-Bloch equation. We show that 
infinitely many conserved local integrals resulting from the integrability 
of the equation can be obtained in a general group theoretic framework of 
symmetric space $G/H$. This enlarges previously known results in the case of 
the two-level system and provides new conserved charges in other multi-level 
cases. More importantly, our field theory reveals new types of symmetries; 
i) global gauge symmetry, ii) global $U(1)$-axial vector symmetry, iii) 
chiral symmetry and iv) dual symmetry. We show that global gauge symmetry can 
be used to generate simulton solutions systematically. Global $U(1)$-axial 
vector symmetry gives rise to the nontopological charge via the Noether method.
Chiral and dual symmetries are discrete symmetries and they 
generate new solutions from a known one. In particular, dual symmetry relates 
the ``bright" soliton with the ``dark" soliton of SIT. 
Finally, we show that the matrix potential is useful in understanding the 
inverse scattering method itself. The potential variable $g$ reveals the 
group structure of the inverse scattering method and we construct explicitly 
soliton solutions for various cases. 
\vglue .2in
The plan of the paper is the following; in Sec. 2, we present a field 
theory formulation of the Maxwell-Bloch equation. 
The area theorem and the sine-Gordon field theory limit are briefly reviewed 
and an extension to the complex sine-Gordon field theory is made in Sec. 2.1. 
In Sec. 2.2, a matrix potential formalism is presented and a general action 
principle is found for the Maxwell-Bloch equation for arbitrary multi-level 
systems. In Sec. 2.3, inhomogeneous broadening is also incoporated into the 
matrix potential formalism. Section 3 deals with explicit examples of various 
multi-level systems. Specific group structures and gauge fixing for each 
systems are identified. In Sec. 4, we explain new features of optical pulses 
in our matrix potential formalism. In Sec. 4.1, topological properties of 
pulses are analyzed through the effective potential energy and its degenerate 
vacua and also by defining topological soliton numbers. In Sec. 4.2, 
nontopological solitons are introduced and interpreted as self-detuned pulses. 
In Sec. 4.3, a new analysis of pulse stability is made in terms of newly found 
nontopological charges. Section 5 deals with symmetries of the system.
Infinitely many conserved charges are constructed systematically for the general 
multi-level systems in Sec. 5.1. Global gauge symmetries are explained in 
Sec. 5.2 and the chiral and the dual symmetries are explained in Sec. 5.3.
Finally, Sec. 6 is a discussion.

\section{Field theory for the Maxwell-Bloch equation }
\setcounter{equation}{0}
The multi-mode optical pulses propagating in a resonant medium along 
the $x$-axis are described by the electric field of the form,
\be
{ \bf E} = \sum_{l=1}^{m}{\cal {\bf E}}_{l}(x, t) \exp i( k_{l}x - 
w_{l}t ) + \mbox{c.c.}
\label{elec} 
\ee
where $k_{l}$ and $w_{l}$ denote the wave number and the frequency 
of each mode and the amplitude vector $ {\cal {\bf E}}_{l} $ is in 
general a complex vector function. The governing equation of propagation 
is the Maxwell equation,
\be
({\pp^{2} \over \pp x^{2}} - {n^{2}\over c^{2}}{\pp^{2} \over 
\pp t^{2}} )
{\bf E} = {4\pi \over c^{2}}{\pp^{2} \over \pp t^{2}}\int dv 
\mbox{ tr } \rho {\bf d} .
\label{maxwell}
\ee
On the right hand side, electric dipole transitions are treated 
semiclassically. ${\bf d}$ is the atom's dipole moment operator and 
the density matrix $\rho $ satisfies the  quantum-mechanical optical 
Bloch equation
\be
i\hbar ( {\pp \over \pp t } + v{\pp \over \pp x})\rho = [(H_{0} -
{\bf E}\cdot {\bf d}) \ , \ \rho  ] \ .
\label{Bloch}
\ee
$H_{0}$ denotes the Hamiltonian of a free atom and $v$ is the 
$x$-component of the velocity of the atoms. In general, we make a 
slowly varying envelope approximation (SVEA) for the Maxwell-Bloch 
system where the amplitudes $ {\cal {\bf E}}_{l} $ vary slowly compared 
to the space and time scales determined by $k_{l}$ and $w_{l}$. Under 
SVEA, the Maxwell-Bloch equation becomes a set of coupled first order 
partial differential equations for the amplitudes $ {\cal {\bf E}}_{l} $ 
and the components of the density matrix. Explicit expressions of the 
Maxwell-Bloch equation for several multi-level cases are given 
in Sec. 3. Thus, the Maxwell-Bloch equation provides an effective, 
semiclassical description of light-matter interaction using the amplitudes 
$ {\cal {\bf E}}_{l} $ as dynamical variables. Unlike the linear case, 
it is quite difficult, if not impossible, to introduce a potential 
variable instead of $ {\cal {\bf E}}_{l} $ due to the nonlinearity of the 
interaction and the approximation involved. Lacking a potential variable 
causes the physical system to be described only by the equation of motion, 
not by an action principle. Consequently, a field theoretic formulation is 
lacking in the problem of pulse propagation. However, when pulses propagate 
in a resonant, nondegenerate two-level atomic medium with inhomogeneous 
broadening, McCall and Hahn have introduced an effective potential-like 
variable, and in terms of which shown that an arbitrary pulse evolves into 
a coherent mode of lossless pulses \cite{McCall}. This phenomenon, known as 
self-induced transparency, is also observed in more general, degenerate 
and/or multi-level atomic media. Specifically, McCall and Hahn have shown 
that when the dimensionless pulse envelope function $E$ is assumed to be 
real and the time area of $2E$,
\be
\theta (x) = 2  \int_{-\infty }^{ \infty } dt  ~ E ,
\label{area}
\ee
is an integer multiple of $2\pi $ ( $2n\pi $ pulse), then the pulse 
propagates without loss of energy. Otherwise, due to inhomogeneous 
broadening the pulse quickly reshapes into a $2n \pi $ pulse  according to 
the area theorem,
\be
 {d \theta (x) \over dx } = -\a \sin{ \theta (x) } 
\label{areathm}
\ee
for some constant $\a $. The proof of the area theorem can be done by 
making use of inhomogeneous broadening and the Maxwell-Bloch equation. 
In the absence of inhomogeneous broadening, the system was shown to 
admit the sine-Gordon field theory formulation.

\subsection{The sine-Gordon limit}
The Maxwell-Bloch equation for the nondegenerate two-level case can be 
written in a dimensionless form by 
\ben
\pb E + 2 \b <P> &=& 0 \nonumber \\
\pp D - E^{*}P - EP^{*} &=& 0 \nonumber \\
\pp P + 2i\xi  P + 2ED &=& 0
\label{sit}
\een
where $\b $ is a coupling constant and $\xi  = w-w_{0} \ , \ \pp \equiv 
\pp /\pp z \ , ~ \ \pb \equiv \pp / \pp \bar{z} \ , z= t-x/c , ~  \bar{z} = x/c$. 
The angular bracket signifies an average over the spectrum distribution 
$f(\xi  )$ as given by
\be
< \cdots > = \int^{\infty }_{- \infty } ( \cdots )f(\xi  )d\xi .
\label{inhomog}
\ee
The dimensionless quantities 
$E,P$ and $ D$ correspond to the electric field, the polarization and 
the population inversion through the relation,
\ben
E &=& -i{\bf E}\cdot{\bf e} t_0 d/\sqrt{6} \hbar \nonumber \\ 
P &=& -\r_{12} \exp [-i(kx-\o t)]/4k t_0 N_0 f(\x ) \nonumber \\ 
D &=& -(\r_{22}-\r_{11})/8 k t_0 N_0 f(\x ) 
\label{epd}
\een
where ${\bf e}$ specifies the linear polarization direction, 
$t_0$ is a time constant and $N_0$ is related to the stationary 
populations of the levels.\footnote{For the details of constants, we 
refer the reader to Ref. \cite{maim2}.} 
In order to understand the structure of $2n \pi $ pulses 
better, we impose further restrictions such that the system is on 
resonance ($\xi  =0$), frequency modulation is ignored ($E$ being 
real) and inhomogeneous broadening is absent ($f(\xi ) = \d (\xi )$). 
Under such restrictions, we could introduce an area function $\vf (x,t)$ 
defined by
\be
\vf (x,t) \equiv \int_{-\infty }^{t} Edt^{'} ,
\label{areaftn}
\ee
which, in the limit $t \rightarrow \infty $, agrees with $\theta(x)/2$ 
in (\ref{area}). In terms of $\vf $, the SIT equation reduces to the 
well-known sine-Gordon equation,
\be
\pb\pp \vf - 2\b \sin{2\vf } = 0 ,
\label{sg}
\ee
when we make consistent identifications;
\be
E=E^{*} = \pp \vf \ \ , \  \ <P> = P = -\sin{2\vf } \ \ , \ \ 
<D> = D= \cos{2\vf } \ .
\label{epd2}
\ee
This sine-Gordon equation arises from the action
\be
S =  {1 \over 2\pi } \int ( \pp \vf \pb \vf - 2\b \cos{2\vf } ) .
\ee
The periodic cosine potential term exhibits infinitely many degenerate 
vacua. It gives rise to soliton solutions which interpolate between two 
different vacua. This shows that the $2n \pi $ pulse can be identified 
with the topological $n$-soliton solution of the sine-Gordon equation.
The electric field amplitude $E$, now identified with 
$\pp \vf $, receives an interpretation as a topological current. 
Note that the area function $\vf $ is different from the 
conventional scalar or vector potentials of electromagnetism. 
Nevertheless, it is remarkable that the potential energy $\cos {2\vf }$ 
of the sine-Gordon Lagrangian can be identified with the population 
inversion $D$ which represents the atomic energy. Also the Lorentz 
invariance, which was broken by SVEA, re-emerges in the 
sine-Gordon field theory after the redefinition of coordinates. 
The identification of the atomic energy with the cosine potential term 
shows that $2n\pi $ pulses are stable against finite energy fluctuations 
due to the conservation of the topological number $n$.
\\

Though the sine-Gordon theory provides a nice field theory for the 
nondegenerate two-level system, it is too restrictive for real 
applications. The presense of frequency modulation in pulses, for example, 
require that $E$ should be complex. Therefore, in this case $E$ can not be 
simply replaced by a real scalar field $\vf $ and the sine-Gordon limit is no 
longer valid. Also, inclusion of frequency modulation invalidates the area 
theorem. However, through the inverse scattering method, it has been 
found that solitons do exist even in the case of complex $E$ \cite{lamb}. This 
suggests that a more general field theory of SIT than the sine-Gordon 
theory could exist which takes care of a complex $E$. Recently, we have 
shown that this is indeed true and the field theory which includes both 
the frequency detuning and the modulation effects is the so-called 
``complex sine-Gordon theory" \cite{park1}. This generalizes the sine-Gordon 
theory as follows; 
assume that $E$ is complex and the frequency distribution function of 
inhomogeneous broadening is sharply peaked at $\xi $, 
i.e. $f(\xi ^{'}) = \d (\xi ^{'}  - \xi )$ for some constant $\xi $.  
Introduce parametrizations of $E, ~P$ and $D$, which generalize parametrizations 
in Eq. (\ref{epd2}), in terms of three scalar fields $\vf , ~ \q $ and $\h $,
\be
E = e^{i(\q - 2\h )}( 2\pp \h {\cos{\vf } \over \sin{\vf }} - 
i\pp \vf )  \ \ , \ \ 
P = ie^{i(\q - 2\h )}\sin{2\vf } 
\ \ , \ \ 
D = \cos{ 2\vf } \ .
\label{csgepd}
\ee
These parametrizations consistently reduce the two-level Maxwell-Bloch 
equation (\ref{sit}) into a couple of second order nonlinear differential 
equations known as the complex sine-Gordon equation;
\ben
\pb\pp \vf + 4{\cos{\vf } \over \sin^{3}{\vf }}\pp \h \pb \h -
2\b \sin{2\vf } & =& 0  
\nonumber \\
\pb \pp \h - {2 \over \sin{2\vf }}(\pb \h \pp \vf + \pp \h \pb \vf 
) &=& 0
\label{csg}
\een
and a couple of first order constraint equations,
\ben
2\cos^{2}{\vf }\pp \h - \sin^{2}{\vf }\pp \q - 2\xi \sin^{2}{\vf } 
&= & 0 \nonumber \\
2\cos^{2}{\vf }\pb \h + \sin^{2}{\vf }\pb \q  &=& 0 \ .
\label{constraint}
\een
Note that the complex sine-Gordon equation reduces to 
the sine-Gordon equation when frequency modulation is ignored 
so that $\h =0, \ \q = \pi / 2 $ and the system is on 
resonance ($\xi = 0$). This reduction is consistent with the original 
equation since solutions of the sine-Gordon equation consists a 
subspace of the whole solution space. The complex sine-Gordon equation 
was first introduced by Lund and Regge in 1976 in order to describe 
the motion of relativistic vortices in a superfluid  \cite{lund}, 
and also independently by Pohlmeyer in a reduction problem of O(4) 
nonlinear sigma model \cite{pohl}. This equation is known to be 
integrable and soliton solutions generalizing those of the sine-Gordon 
equation have been found.  These issues will be considered in later 
sections in a more general context. The Lagrangian for the complex 
sine-Gordon equation in terms of $\vf $ and $\h $ is given by
\be
S = {1 \over 2\pi }\int \pp \vf \pb \vf + 4 \cot^{2}{\vf } \pp \h \pb \h 
- 2\b \cos {2\vf } .
\label{csglag}
\ee
This Lagrangian, however, is singular at $\vf = n \pi $ for integer $n$ 
which causes difficulties in quantizing the theory. Also, besides the 
complex sine-Gordon equation, the two-level Maxwell-Bloch equation comprises 
the constraint equation (\ref{constraint}). Thus the Lagrangian (\ref{csglag}) 
does not quite serve for a field theory action of the two-level system. 
In fact, the singular behavior of the Lagrangian (\ref{csglag}) is 
an artifact of neglecting the constraint equation. This, as well as 
the rationale of the above parametrizations, can be seen most clearly if 
we reformulate the Lagrangian to include the constraint in the context 
of a matrix potential and a gauged nonlinear sigma model as explained in 
the next section.

\subsection{Matrix potential formalism }

In order to construct a field theory action of the Maxwell-Bloch equation in 
terms of potential variables and also find a way to extend to more 
general multi-level and degenerate cases, we first note that the optical 
Bloch equation admits an interpretation of a spinning top equation as in the 
case of the corresponding magnetic resonance equations \cite{bloch}. Denote 
real and imaginary parts of $E$ and $P$ by $E = E_{R} + iE_{I}, ~ P = 
P_{R} + iP_{I}$. Then, the Bloch equation in Eq. (\ref{sit}) 
can be expressed as 
\be
\pp \vec{S} = \vec{\Omega } \times \vec{S}
\label{topeq}
\ee
where $\vec{S} = (P_{R}, ~ P_{I}, ~ D), ~~ \vec{\Omega } = 
(2E_{I}, ~ -2E_{R}, ~-2\xi )$. This describes a spinning top 
where the electric dipole ``pseudospin" vector $\vec{S}$ precesses 
about the ``torque" vector $\vec{\Omega }$. This clearly shows that 
the length of the vector $\vec{S}$ is preserved,
\be
|\vec{S}|^{2} = P_{R}^{2} + P_{I}^{2} + D^{2} = 1.
\label{prob}
\ee
The length equals unity due to the conservation of probability.
The remaining Maxwell equation in Eq. (\ref{sit}) determines the strength 
of the torque vector. If $P_{I} = 0$, we may solve (\ref{prob}) by taking 
$P_{R} = -\sin{2\vf } $ and $D = \cos{2\vf }$ and also (\ref{topeq}) 
by taking $E = \pp \vf $ as given in (\ref{epd2}). Then, the Maxwell 
equation becomes the sine-Gordon equation as before. This picture 
agrees with the conventional interpretation of the sine-Gordon theory 
as describing a system of an infinite chain of pendulums.
In order to generalize the sine-Gordon limit to the complex $E$ and 
$P$ case, we make a crucial observation that the constraint in 
Eq. (\ref{prob}) can be solved in general in terms of an $SU(2)$ matrix 
potential variable $g$ by
\be
\pmatrix{D & P \cr P^{*} & -D} = g^{-1}\s_{3}g, 
\label{gspin}
\ee
where $\s_{3} = \mbox{diag}(1, -1)$ is the Pauli spin matrix. By taking 
the determinent, one can check that Eq. (\ref{prob}) is 
automatically satisfied. Also, note that $g^{-1}\s_{3}g$ 
is invariant under the ``chiral $U(1)$-transformation'', 
\be
g \rightarrow e^{i f \s_{3}}g
\label{uonesym}
\ee
for any function $f$. Thus, $g^{-1}\s_{3}g$ parameterizes $SU(2)/U(1)$ 
instead of $SU(2)$. Since Eq. (\ref{prob}) is automatically solved in 
this $SU(2)$ parametrization, the number of independent variables, which 
is $\mbox{dim}(SU(2)) - \mbox{dim}(U(1)) = 3-1=2$, agrees with that of 
the vector ($\vec{S}$) parametrization. Moreover, we have an identity
\be
\pp (g^{-1} \s_{3} g) = [ g^{-1}\s_{3}g, ~ g^{-1} \pp g] ,
\ee                                                       
where the bracket denotes a commutator. Note that this becomes precisely 
the Bloch equation if we make an identification,
\be
g^{-1} \pp g + R =  \pmatrix{ i\xi  & -E \cr E^{*} & -i\xi } ,
\label{idelec}
\ee 
where $R$ is an anti-Hermitian matrix commuting with $g^{-1}\s_{3}g$ 
which will be determined later. Thus, we have solved the Bloch equation 
through the matrix potential $g$ up to the identification in 
Eq. (\ref{idelec}). The identification is consistent since both sides are 
anti-Hermitian matrices. The off-diagonal part of the r.h.s. is simply 
renaming the component variable by $E$ whereas the constant diagonal 
part imposes a constraint on the variable $g$. This constraint, however, 
can be satisfied by an appropriate chiral $U(1)$-transformation in 
Eq. (\ref{uonesym}). Thus, the matrix potential $g$ is made of two 
independent variables and one variable satisfying the constraint. 
In the following, we show that the Maxwell equation can be expressed 
in terms of two independent variables only, decoupling completely from 
the constraint variable. In this regard, our matrix potential $g$ 
resembles the scalar potential $\vf$ in electrostatics where $\vf $ 
solves the curl-free condition, $\nabla \times \vec{E} = 0$, 
identically and changes the Gauss equation into the Poisson equation. 
In our case, the Schr\"{o}dinger equation plays the role of the 
curl-free condition and the Maxwell equation, the counterpart of the 
Gauss equation, changes into a second order nonlinear differential 
equation. In order to see this, observe that the Maxwell equation 
can be expressed also in terms of $g$ only,
\be
\pb (g^{-1} \pp g + R) = \pmatrix{ 0 & -\pb E \cr \pb E^{*} & 0 } 
=  \b[ \s_{3}, ~ g^{-1}\s_{3} g].
\label{maxpot}
\ee
Thus, we have successfully expressed the SIT equation in terms of the 
potential variable $g$ up to an undetermined quantity $R$.
As we will show, $R$ is determined by requiring an action principle 
for the Maxwell equation in terms of $g$. Since $g$ is constrained, 
we need a Lagrange multiplier for the constraint. In order to help 
understanding, we assume for a moment that $R = 0$ and the system 
is on resonance ($\xi = 0$). Then, the equation of motion 
(\ref{maxpot}) arises from a variation of the action
\be
S = S_{WZNW}(g)  - S_{\mbox{pot}} + S_{\mbox{const}}
\label{act}
\ee 
with the following variational behaviors;
\ben
\d _{g}S_{WZNW} &=&  {1 \over 2\pi }\int dz d\zb \mbox{Tr} (
\pb ( \gi \pp g ) \gi \d g )
\nonumber 
\\ 
\d_{g}S_{\mbox{pot}} &=& {\b \over 2\pi }\int dz d\zb \mbox{Tr}(
 [ \s_{3}, ~ \gi \s_{3} g \ ]\gi \d g ).
\label{zeroeq}
\een
The action $S_{WZNW}(g)$ is the well-known $SU(2)$ 
Wess-Zumino-Novikov-Witten functional,
\be
S_{WZNW}(g)=-{1\over 4\pi }\int_{\S }dz d\zb 
\mbox{Tr }( \gi \pp g \gi \pb g) - {1 \over 12\pi }\int_{B}\mbox{Tr } 
(\tilde{g}^{-1}d \tilde{g}\wedge \tilde{g}^{-1}d \tilde{g} \wedge 
\tilde{g}^{-1}d \tilde{g}) \ ,
\label{wzw}
\ee
where the second term on the r.h.s., known as the 
Wess-Zumino term, is defined on a three-dimensional manifold $B$ with 
boundary $\S = \pp B $ and $\tilde{g}$ is an extension of a map $g:\S 
\rightarrow SU(2)$ to $B$ with $ \tilde{g}|_{ \S }=g$ \cite{Witten}.
The potential term $S_{\mbox{pot}}$ can be easily obtained by
\be
S_{\mbox{pot}}= {\b \over 2\pi }\int dz d\zb \mbox{Tr}(g\s_{3} \gi 
\s_{3}) .
\label{potential}
\ee
Finally, the constraint requires vanishing of the diagonal part of the 
matrix $g^{-1}\pp g$ which can be imposed by adding a Lagrange 
multiplier term $S_{\mbox{const}}$ to the action
\be
S_{\mbox{const}} = {1 \over 2\pi }\int  dz d\zb 
\mbox{Tr}( \l \s_{3} g^{-1}\pp g ).
\ee
The Lagrange multiplier $\l $, however, induces a new term to the equation 
of motion by
\be
\d_{g}S_{\mbox{const}} = {1 \over 2\pi }\int dz d\zb \mbox{Tr}((-\pp 
\l\s_{3} + [\l \s_{3}, ~ g^{-1}\pp g ])g^{-1}\d g) ,
\ee
which seems to spoil our construction of a  field theory. 
This problem can be resolved beautifully if we introduce a gauge symmetry 
and  make the action (\ref{act}) to be ``vector gauge invariant". 
This can be done by replacing the constraint term with a ``gauging" 
part of the Wess-Zumino-Novikov-Witten action,
\ben
S &=& S_{WZW}(g)  - S_{\mbox{pot}} + S_{\mbox{gauge}} 
\label{action} \\
S_{\mbox{gauge}} &=& 
{1 \over 2\pi }\int \mbox {Tr} (- A\pb g \gi + \Ab \gi \pp g
 + Ag\Ab \gi - A\Ab )
\een
where the connection fields $A, \Ab$ gauge the anomaly free subgroup 
$U(1)$ of $SU(2)$ generated by the Pauli matrix $\s_{3}$. They 
introduce a $U(1)$-vector gauge invariance of the action where the 
$U(1)$-vector gauge transformation is defined by\footnote{ Note that 
here we are using the vector $U(1)$-transformation instead of the 
chiral one as in Eq. (\ref{uonesym}). This causes the matrix 
$g^{-1}\s_{3} g$ to transform covariantly under the gauge 
transformation.}
\be
g \rightarrow h^{-1}gh \ \ , \ \ A \rightarrow h^{-1}Ah + 
h^{-1}\pp h \ \ , \ \  \Ab \rightarrow h^{-1}\Ab h  + h^{-1}\pb h 
\label{gaugetr}
\ee
where $h = \exp( i f\s_{3} )$ for some scalar function $f$. 
Owing to the absence of kinetic terms, $A, \Ab $ act as Lagrange 
multipliers which result in the constraint equations when $A$ and 
$ \Ab $ are integrated out. The action in Eq. (\ref{act}) may be 
understood as a gauge fixed action with the choice of gauge where 
$A = 0, ~ \Ab = \l \s_{3}$. The main reason for introducing 
a gauge transformation and a gauge invariant action is twofold. 
It first shows that the equation of motion resulting from the 
variation $\d_{g} S =0$, 
\be
\pb ( \gi \pp g + \gi A g ) + [\Ab , ~ \gi \pp g + \gi A g ] 
- \pp \Ab = \b [\s_{3} , ~ \gi \s_{3} g ] ,
\label{ginvmax}
\ee
is also gauge invariant and gives rise to the gauge invariant 
expression of the Maxwell equation. Comparing Eq. (\ref{ginvmax}) 
with Eq. (\ref{maxpot}), we see that $R=\gi A g$ and 
the relevant gauge choice is $A= i\xi \s_{3} , ~ \Ab =0$ due to the 
constraint in Eq. (\ref{const3}).\footnote{One 
can always choose such a gauge due to the
flatness of $A$ and $\Ab$ as in Eq. (\ref{flat}).}  The 
$U(1)$-vector gauge invariance of the Maxwell equation implies 
that the Maxwell equation is independent of specific gauge 
choices. Thus, it decouples from the $U(1)$ scalar field which 
saturates the constraint condition and becomes a couple of 
second order nonlinear differential equations in two local 
variables. In the next section, this is shown clearly by an 
explicit parametrization of $g$ and the resulting Maxwell equation 
is shown to be equivalent to the complex sine-Gordon equation given 
in Eq. (\ref{csg}). The second reason is that gauge transformation 
incoporates beautifully the frequency detuning effect through 
specific gauge fixing. In Sec. 3, external magnetic fields are 
also incoporated through gauge fixing.
\vglue .2in
Our field theory for the Maxwell-Bloch equation is not restricted 
to the two-level case. In fact, the group theoretical formulation 
through the matrix potential $g$ allows an immediate extension to 
the multi-level cases. We may simply replace the pair $SU(2) 
\supset U(1) $ by $G \supset H $ for any Lie groups $G$ and $H$ and 
obtain the $G/H$-gauged Wess-Zumino-Novikov-Witten action ($S_{WZNW} + 
S_{\mbox{gauge}}$) where $A$ and $\Ab$ gauge the subgroup $H$ of 
$G$.\footnote{This action is known to possess conformal 
symmetry and has been  used for the general $G/H$-coset conformal 
field theories \cite{coset}. The potential energy term 
(\ref{potential}) breaks conformal symmetry. Nevertheless, it 
preserves the integrability of the model given by (\ref{action}) 
where $G/H = SU(2)/U(1)$, and this model has been used in describing 
integrable perturbation of parafermionic coset conformal field theories 
\cite{bakas,park}.} 
For a general pair of $G$ and $H$, the expression for the potential 
which preserves integrability can be given by \cite{park,shin2},
\be
S_{\mbox{pot}}= {\b \over 2\pi }\int \mbox{Tr}(gT\gi \Tb )
\label{potent}
\ee
where $T$ and $ \Tb $ are constant matrices which commute with the 
subgroup $H$, i.e. $[T, h] = [\Tb , h ] = 0, \mbox{ for } h \in H$.
This makes the potential term vector gauge invariant.\footnote{ 
In a more general context, $S_{\mbox{pot}}$ is specified 
algebraically by a triplet of Lie groups $F \supset G \supset H$ 
for every symmetric space $F/G$, where the Lie algebra decomposition 
${\bf f} = {\bf g} \oplus {\bf k} $ satisfies the commutation 
relations,
\be
[{\bf g} ~ , ~ {\bf g}] \subset {\bf g} ~ , ~ [ {\bf g} ~ , ~ {\bf k}] 
\subset {\bf k} ~ , ~ [{\bf k} ~ , ~ {\bf k} ] \subset {\bf g} ~ .
\label{algebra}
\ee
We take $T$ and $\Tb$ as elements of $ {\bf k}$ and  define  
$ {\bf h} $ as the simultaneous centralizer of $T$ and $\Tb $, i.e. 
${\bf h} = C_{{\bf g}}(T, \Tb ) = \{ B \in {\bf g} \ : \ [B ~ , ~ T] = 
0 = [B ~ , ~ \Tb]\} $ with $H$ its associated Lie group. 
With these specifications, the action (\ref{action}) becomes integrable 
and generalizes the sine-Gordon model according to each symmetric 
spaces. For compact symmetric spaces of type II, e.g. symmetric spaces 
of the form $G \times G /G$, the model becomes equivalent to the 
type I case but with $T $ and $\Tb$ belonging to the Lie algebra 
${\bf g}$. It has the coset structure $G/H$ where $H$ is the stability 
subgroup of $T, ~ \Tb \in {\bf g}$ \cite{bps}.  }
As we will see later, physically interesting cases all correspond to 
a special type of symmetric spaces $G/H$, known as Hermitian symmetric 
spaces \cite{helgason}, where the adjoint action of $T$ defines a complex 
structure on $G/H$. 
\\

Now, we define the field theory action for the Maxwell-Bloch 
equation by 
\be
S_{MB} = S_{WZNW}(g)  + 
{1 \over 2\pi }\int \mbox {Tr} (- A\pb g \gi + \Ab \gi \pp g
 + Ag\Ab \gi - A\Ab ) - 
 {\b \over 2\pi }\int \mbox{Tr}(gT\gi \Tb ) .
\label{action2}
\ee
This action is of course restricted to the integrable cases which 
require specific fine tuning of coupling constants. However, the 
concept of matrix potential $g$ is valid irrespective of the 
integrability of the model and the field theory formulation can be 
extended to more general, nonintegrable cases too. In this paper, 
we will restrict only to the integrable cases.
The equation of motion arising from the variation of the 
action (\ref{action2}) with respect to $g$ gives rise to the Maxwell 
equation in the matrix potential formalism,
\ben                     
\underline{Maxwell ~eq.}: ~~~~~~~~~~~~~~~~~~~~~~~~~~~~~~~~~~~~~~~~~~~~~
~~~~~~~~~~~~~~~~~~~~~~~~~~~~~~~~~~~~~~~~
 \nonumber \\
 \pb (\gi \pp g + \gi A g )  + [\Ab , ~ \gi \pp g + \gi A g ]- \pp \Ab 
= \b [T, ~ \gi \Tb g ] .~~~~~~~~~~~~~~    
\label{zeroeqn}
\een 
The Bloch equation again arises from the simple identity
\ben
\underline{Bloch ~eq.}: & \nonumber \\
& \pp ( \gi \Tb g ) = [\gi \Tb g , ~  \gi \pp g + \gi A g ] ,~~~~~~~~~~~~~~~~~~~~~~~~
~~~~~~~~~~~~~~~
\een 
where we used the property $[\Tb , ~ A] = 0$.
This rather abstract form of the Maxwell-Bloch equation will become  
more explicit when specific identifications of physical variables are 
made in Sec. 3. Note that the Maxwell equation is invariant under 
the $H$-vector gauge transformation as given in Eq. (\ref{gaugetr}), where 
the local function $h$ now belongs to the subgroup $H$, while the 
Bloch equation is not. The integrability of the Maxwell equation may 
be demonstrated by rewriting Eq. (\ref{zeroeqn}) in an equivalent zero 
curvature form in terms of the $U-V$ pair, 
\be                                                                
[\pp - U, ~ \pb -V]=0,
\label{uvpair}
\ee
and
\ben
U &\equiv&  - \gi \pp g - \gi A g - \b\l T 
\nonumber \\
V &\equiv& - \Ab - {1 \over \l }\gi \Tb g . 
\label{uv}
\een
Here, $\l $ is an arbitrary spectral parameter. This shows that the 
equation of motion becomes the integrability condition of the overdetermined 
linear equations;
\ben
(\pp - U)\Psi &=& 
(\pp + \gi \pp g + \gi A g + \b\l T  ) \Psi = 0  
\nonumber \\
(\pb - V)\Psi &=& ( \pb + \Ab + {1 \over \l } \gi \Tb g  ) \Psi = 0 .
\label{lineareqn}
\een
The constraint equations coming from the variation of $S_{MB}$ 
with respect to $A, \Ab $ are
\ben
\d _{A}S_{MB} &=& {1 \over 2\pi }\int \mbox{Tr}( ( \ - \pb g 
\gi + g\Ab \gi - \Ab \  )\d A )
= 0  \nonumber \\
\d _{\Ab }S_{MB} &=& {1 \over 2\pi }\int \mbox{Tr}( ( \  \gi 
\pp g  +\gi A g - A \ )\d\Ab ) = 0 \ .
\label{constraint2}
\een
Or,
\be
( \ - \pb g \gi + g\Ab \gi - \Ab \  )_{\bf h} = 0 , ~~ 
( \  \gi \pp g  +\gi A g - A )_{\bf h} = 0  
\label{const3}
\ee
where the subscript ${\bf h}$ specifies the projection to the 
subalgebra ${\bf h}$. It can be readily checked that these constraint 
equations, when combined with Eq. (\ref{zeroeqn}), imply the flatness of 
the connection $A$ and $ \Ab $, i.e.
\be
F_{z \zb } = [ \ \pp + A \ , \ \pb + \Ab \ ] = 0 \ .
\label{flat}
\ee
In Sec. 3, we show that various multi-level Maxwell-Bloch equations 
indeed arise from Eqs. (\ref{zeroeqn}) and (\ref{const3}) when  
appropriate choices are made for the groups $G$ and $H$, the constant matrices 
$T$ and $ \Tb $, and gauge fixing. 
\subsection{Inhomogeneous broadening}
So far, we have obtained an action principle for the Maxwell-Bloch 
equation without inhomogeneous broadening. Remarkably, even in the 
presense of inhomogeneous broadening, the notion of matrix potential 
still persists. The inhomogeneous broadening effect, i.e. Doppler 
shifted atomic motions, can be incorporated beautifully via the $U(1)$ 
vector gauge transformation. Due to the microscopic motion of atoms, 
each atom in a resonant medium responds to the macroscopic 
incoming light with different Doppler shifts of transition frequencies. 
Thus, microscopic variables, e.g. the polarization $P$ and the 
population inversion $D$ are characterized by Doppler shifts and they 
couple to the macroscopic variable $E$ through an average over 
the frequency spectrum as given in (\ref{inhomog}). A remarkable 
property of our effective field theory formulation is that it 
includes inhomogeneous broadening naturally only with minor 
modifications. The notion of potential variable $g$ is again valid.  
In order to cope with microscopic motions, $g$ becomes a 
function of frequency $\xi $, i.e. $g = g(z, \zb , \xi )$. 
However, the action principle in (\ref{action2}) is no longer valid 
despite the use of the potential variable $g$. We also 
relax the constraint in Eq. (\ref{constraint2}) and require only
\be
(\gi \pp g   +\gi A g)_{\bf h} - A  = 0 \ .
\label{inhocon}
\ee
Then, the linear equation is given by
\ben
L_{z}\Psi &\equiv & \left( \partial + g^{-1}\partial g +  
g^{-1}Ag - \xi T + \tilde{\lambda }T \right) \Psi = 0 \nonumber \\
L_{\zb }\Psi &\equiv &
\left( \ \bar{\partial} + 
\left< {g^{-1}\bar{T} g \over \tilde{\lambda } - \xi ^{'} 
} \right> \right) \Psi = 0
\label{inholin}
\een
where the constant $\tilde{\lambda }$ is a modified spectral parameter 
and becomes $\lambda + \xi $ in the absense of inhomogeneous 
broadening. The angular brackets denote an average over $\xi^{'}$ as 
in Eq. (\ref{inhomog}). As in the case without inhomogeneous 
broadening, we make the same identification of the matrix 
$g^{-1}\partial g +  g^{-1}Ag - \xi T $ with various components of 
macroscopic electric fields which are independent of the microscopic 
quantity $\xi$. This requires the $\xi$-dependence of 
$g(z, \zb , \xi )$ to be determined in such a way that 
$g^{-1}\partial g +  g^{-1}Ag - \xi T $ is independent of $\xi $. 
It is easy to see that this requirement is indeed satisfied by various 
integrable Maxwell-Bloch systems considered in Sec. 3.\footnote{ In the 
three-level system, we must take $\x = -t_0 \D_1 = -t_0 \D_2$. It means 
that in order to preserve the integrability in the presence of inhomogenuous 
broadening, two detuning parameters of the three-level system must be 
equal.}  Note that $\Psi (\tilde{\l} , z, \zb )$ is not a function of 
$\x$. The integrability of the linear equation (\ref{inholin}) becomes
\ben
0 &=& \left[ \partial + g^{-1}\partial g + g^{-1}Ag - \xi T + 
\tilde{\lambda }T \ , \ \bar{\partial} + 
\left< {g^{-1}\bar{T} g \over \tilde{\lambda } - \xi^{'} } \right> 
\right] 
\nonumber \\
&=& - \pb ( g^{-1}\partial g +  g^{-1}Ag - \xi T ) 
+ \left< [T, ~ g^{-1}\bar{T} g ] \right>
\label{inhozero}
\een
where we used the fact that $g^{-1}\partial g + g^{-1}Ag - \xi T $ is 
independent of $\xi $ and also the identity
\be
\pp ( g^{-1}\bar{T} g ) + [ g^{-1}\partial g + g^{-1}Ag , ~ 
g^{-1}\bar{T} g ] = 0.
\label{ident}
\ee
Once again, identifying $g^{-1}\bar{T} g$ with components of the density 
matrix and Eq. (\ref{ident}) with the Bloch equation, we obtain the 
Maxwell-Bloch equation with inhomogenous broadening. For example, we may 
identify $E, P$ and $D$ as in Eq. (\ref{epdtwo}) so that Eqs. (\ref{inhozero}) 
and (\ref{ident}) become the Maxwell-Bloch equation with inhomogeneous 
broadening for the nondegenerate two-level case as given in Eq. (\ref{sit}). 
Note that each frequency $\xi $ corresponds to a specific gauge choice of the 
vector $U(1)$ subgroup. Therefore, in some sense inhomogenous broadening is 
equivalent to averaging over different gauge fixings of $U(1) \subset H$. 
This implies that inhomogenous broadening can not be treated by a single 
field theory and therefore it lacks a Lagrangian formulation. It is 
remarkable, however, that the group theoretic parametrization of various 
physical variables in terms of the potential $g$ still survives.
 
\section{Multi-level systems}      
\setcounter{equation}{0}
In this section,  we work out in detail field theory identifications
of each multi-level Maxwell-Bloch equations through specifying the groups 
$G$ and $H$, the constant matrices $T$ and $ \Tb $, and the gauge choice. 
Briefly, the resulting associations with symmetric spaces are the 
following (see Fig. 1a-1h); 
\ben
SU(2)/U(1) & \leftrightarrow & \mbox{nondegenerate two-level ~ (Fig.1a)}
\nonumber \\
SU(3)/U(2) & \leftrightarrow & \mbox{degenerate two-level;  (Fig.1b and Fig.1c)} 
\nonumber \\
&& j_{b} = 0  \rightarrow j_{a} = 1, ~ j_{b} = 1 \rightarrow j_{a} = 0, ~
j_{b} = 1 \rightarrow j_{a} = 1 \nonumber \\
(SU(2)/U(1))^{2} & \leftrightarrow & \mbox{degenerate two-level};
j_{b} =1/2 \rightarrow j_{a} = 1/2 ~ \mbox{(Fig.1d)} \nonumber \\
SU(3)/U(2) & \leftrightarrow & \mbox{nondegenerate three-level, 
$\L$  or $V$  system (Fig.1e and Fig.1f)} \nonumber \\
SU(4)/S(U(2) \times U(2)) & \leftrightarrow & 
\mbox{degenerate three level};
~ j_{a} = j_{c} =  0, ~ j_{b} = 1 ~ \mbox{(Fig.1g)}.
\nonumber \\
SU(5)/U(4) & \leftrightarrow & \mbox{degenerate three level};
~ j_{a} = j_{c} = 1, ~ j_{b} =0 ~ \mbox{(Fig.1h)}. \nonumber \\
&&   
\label{exherm}
\een

\begin{figure}                                       
               
\unitlength 1cm
\begin{picture}(14,15)(1,-15)
\put(1,-1.6){\begin{picture}(4,5)(0,0.5)
\put(0.5,0.5){\line(1,0){1}}
\put(0.5,2.5){\line(1,0){1}}
\put(1,0.5){\vector(0,1){2}}
\put(1,2.5){\vector(0,-1){2}}
\put(1.05,1.45){$E$}
\put(-0.2,-0.5){(a) $SU(2)/U(1)$}
\end{picture}}

\put(5,-1.6){\begin{picture}(4,5)(0,0)
\put(0,0){\line(1,0){1}}
\put(2,0){\line(1,0){1}}
\put(1,2){\line(1,0){1}}
\put(0.5,0){\vector(1,2){1}}
\put(1.5,2){\vector(-1,-2){1}}
\put(1.5,2){\vector(1,-2){1}}
\put(2.5,0){\vector(-1,2){1}}
\put(0.3,0.9){$\e^{-1}$}
\put(2.3,0.9){$\e^{1}$}
\put(2.1,1.95){$j_b =0,1$}
\put(1.3,2.25){$m_b =0$}
\put(3.1,-0.1){$j_a =1$}
\put(-0.2,-0.4){$m_a =-1$}
\put(1.8,-0.4){$m_a =1$}
\put(0.5,-1){(b) $SU(3)/U(2)$}
\end{picture}}

\put(11,-1.6){\begin{picture}(4,5)(0,0)
\put(0,2){\line(1,0){1}}
\put(2,2){\line(1,0){1}}
\put(1,0){\line(1,0){1}}
\put(0.5,2){\vector(1,-2){1}}
\put(1.5,0){\vector(-1,2){1}}
\put(1.5,0){\vector(1,2){1}}
\put(2.5,2){\vector(-1,-2){1}}
\put(0.3,0.9){$\e^{-1}$}
\put(2.3,0.9){$\e^{1}$}
\put(3.1,1.95){$j_b =1$}
\put(2.1,-0.1){$j_a =0,1~ m_a = 0$}
\put(-0.2,2.1){$m_b =-1$}
\put(1.8,2.1){$m_b =1$}
\put(0.5,-1){(c) $SU(3)/U(2)$}
\end{picture}}

\put(1,-5.8){\begin{picture}(7,5)(0,0)
\put(0,0){\line(1,0){1}}
\put(2,0){\line(1,0){1}}
\put(0,2){\line(1,0){1}}
\put(2,2){\line(1,0){1}}
\put(0.5,0){\vector(1,1){2}}
\put(2.5,2){\vector(-1,-1){2}}
\put(2.5,0){\vector(-1,1){2}}
\put(0.5,2){\vector(1,-1){2}}
\put(0.3,0.5){$\e^{1}$}
\put(2.3,0.5){$\e^{-1}$}
\put(3.1,1.95){$j_b ={1 \over 2}$}
\put(3.1,-0.1){$j_a ={1 \over 2}$}
\put(-0.2,2.1){$m_b =-{1 \over 2}$}
\put(1.8,2.1){$m_b ={1 \over 2}$}
\put(-0.2,-0.4){$m_a =-{1 \over 2}$}
\put(1.8,-0.4){$m_a ={1 \over 2}$}
\put(0.3,-1){(d) $[SU(2)/U(1)]^2$}
\end{picture}}

\put(6,-5.8){\begin{picture}(4,5)(0,0)
\put(0,0){\line(1,0){1}}
\put(2,0.5){\line(1,0){1}}
\put(1,2){\line(1,0){1}}
\put(0.5,0){\vector(1,2){1}}
\put(1.5,2){\vector(-1,-2){1}}
\put(1.5,2){\vector(2,-3){1}}
\put(2.5,0.5){\vector(-2,3){1}}
\put(0.3,0.9){$\O_2$}
\put(2.3,1.2){$\O_1$}
\put(2.1,1.95){$3$}
\put(3.1,0.4){$1$}
\put(1.1,-0.1){$2$}
\put(0.3,-1){(e) $SU(3)/U(2)$}
\end{picture}}

\put(11, -5.8){\begin{picture}(4,5)(0,0)
\put(0,2){\line(1,0){1}}
\put(2,1.5){\line(1,0){1}}
\put(1,0){\line(1,0){1}}
\put(0.5,2){\vector(1,-2){1}}
\put(1.5,0){\vector(-1,2){1}}
\put(1.5,0){\vector(2,3){1}}
\put(2.5,1.5){\vector(-2,-3){1}}
\put(0.3,0.9){$\O_1$}
\put(2.3,0.6){$\O_2$}
\put(3.1,1.45){$2$}
\put(2.1,-0.1){$3$}
\put(1.1,1.9){$1$}
\put(0.5,-1){(f) $SU(3)/U(2)$}
\end{picture}}

\put(1,-10){\begin{picture}(4,5)(0,0)
\put(0,0){\line(1,0){1}}
\put(2,0){\line(1,0){1}}
\put(0,2){\line(1,0){1}}
\put(3,1.5){\line(1,0){1}}
\put(0.5,0){\vector(2,1){3}}
\put(3.5,1.5){\vector(-2,-1){3}}
\put(0.5,0){\vector(0,1){2}}
\put(0.5,2){\vector(0,-1){2}}
\put(2.5,0){\vector(-1,1){2}}
\put(0.5,2){\vector(1,-1){2}}
\put(2.5,0){\vector(2,3){1}}
\put(3.5,1.5){\vector(-2,-3){1}}
\put(0,0.9){$\e^{1}_1$}
\put(3,0.9){$\e^{-1}_2$}
\put(1.3,1.1){$\e^{-1}_1$}
\put(2.3,1.1){$\e^{1}_2$}
\put(4.1,1.45){$j_c ={0}$}
\put(1.1,1.95){$j_a ={0}$}
\put(3.1,-0.1){$j_b ={1}$}
\put(-0.2,-0.4){$m_b =-{1}$}
\put(1.8,-0.4){$m_b ={1}$}
\put(0.1,-1){(g) $SU(4)/S(U(2) \times U(2))$}
\end{picture}}

\put(8,-10){\begin{picture}(4,5)(0,0)
\put(0,2){\line(1,0){1}}
\put(1.5,2){\line(1,0){1}}
\put(3,1.5){\line(1,0){1}}
\put(4.5,1.5){\line(1,0){1}}
\put(2,0){\line(1,0){1}}
\put(2.5,0){\vector(-1,1){2}}
\put(0.5,2){\vector(1,-1){2}}
\put(2.5,0){\vector(-1,4){0.5}}
\put(2,2){\vector(1,-4){0.5}}
\put(2.5,0){\vector(2,3){1}}
\put(3.5,1.5){\vector(-2,-3){1}}
\put(2.5,0){\vector(4,3){2}}
\put(4.5,1.5){\vector(-4,-3){2}}
\put(0.7,0.9){$\e_1$}
\put(2.3,0.9){$\e_2$}
\put(3.3,0.9){$\e_3$}
\put(4,0.9){$\e_4$}
\put(5.6,1.4){$j_c ={1}$}
\put(-1.1,1.9){$j_a ={1}$}
\put(3.1,-0.1){$j_b ={1}$}
\put(0.2,2.1){$-{1}$}
\put(1.3,2.1){$m_a ={1}$}
\put(2.8,1.6){$m_c =-{1}$}
\put(5,1.6){${1}$}
\put(1.5,-1){(h) $SU(5)/U(4)$}
\end{picture}}

\end{picture}
\vspace{-4.5cm}
\caption{Multi-level systems and their associated symmetric spaces}
\end{figure}
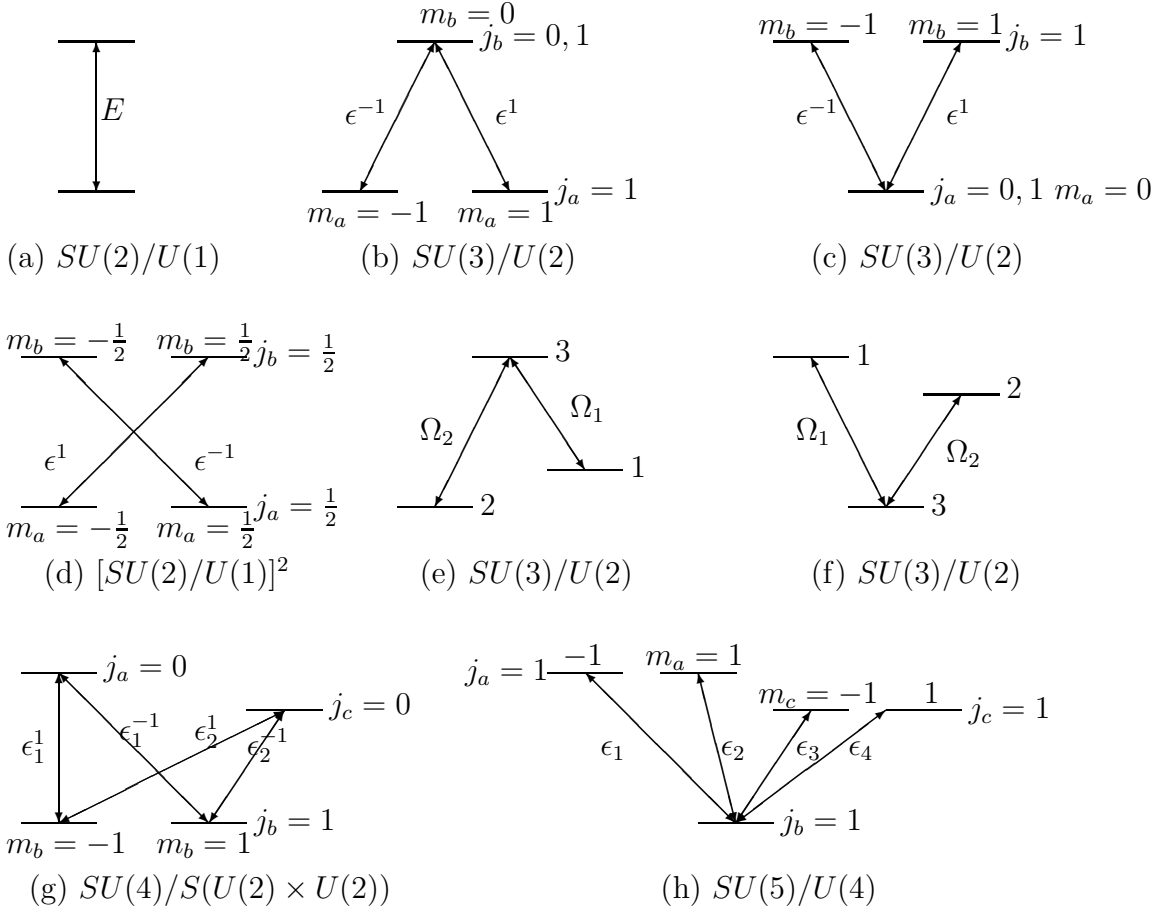

All of them correspond to Hermitian symmetric spaces. In the Appendix B,
the characteristic properties of Hermitian symmetric space is used 
to generate infinitely many conserved local integrals. Our examples in 
Eq. (\ref{exherm}) suggest that to each Hermitian symmetric space there 
may exist a specific multi-level system with a proper adjustment of 
physical parameters. In particular, we could see that the multi-frequency 
generalization in a configuration of the ``bouquet" type \cite{maim2} 
corresponds to the Hermitian symmetric space $SU(n)/U(n-1)$ for an 
integer $n$. However, for large $n$, it requires a fine tuning of 
many coupling constants which makes the theory unrealistic.

\subsection{Nondegenerate two-level system}
This is the simplest case  which was originally considered by 
McCall and Hahn to explain the phenomena of self-induced transparency. 
It also accounts for the transitions $1/2 \rightarrow 
1/2, ~ 1\leftrightarrow 0, ~ 1 \rightarrow 1 $ and $3/2 \leftrightarrow 
1/2$ for linearly polarized waves and the transitions $1/2 \rightarrow 
1/2, ~ 1 \leftrightarrow 0 $ and $ 1 \rightarrow 1$ for circularly 
polarized waves. We assume that inhomogeneous broadening is absent 
so that $<P>=P$. 
The Maxwell-Bloch equation is given by Eq. (\ref{sit}) 
which can be expressed in an equivalent zero cuvature form,
\be
\left[ \  \pp + \pmatrix{ i\b \l + i\xi & -E \cr E^{*} & -i\b 
\l -i\xi } \ , \ \pb - {i \over \l } \pmatrix{D & P \cr P^{*} & -D }
\right]  = 0 \ .
\label{sitzero}
\ee
In order to show that this  equation arises from the field theory 
action in Eq. (\ref{action2}), we take $H= U(1) \subset  
SU(2) = G$ and $ T = - \Tb = i\s _{3} = \mbox{diag}( i , -i )$. 
We fix the vector gauge invariance by choosing 
\be 
A = i\xi \s_{3} \ , \ \Ab =0
\label{gfix}
\ee
for a constant $\xi $. Such a gauge fixing is possible due to the 
flatness of $A, \Ab$. Comparing Eq. (\ref{uv}) with Eq. (\ref{sitzero}), we 
could identify  $E, P$ and $D$ in terms of $g$ such that
\be
g^{-1}\pp g + i \xi  g^{-1}\s_{3} g - i\xi \s_{3} = 
\pmatrix{ 0 & -E  \cr E^{*} & 0 } \ \ , \ \ g^{-1}\s_{3} g = 
 \pmatrix{ D & P \cr P^{*} & -D }
\label{epdtwo}
\ee
which are consistent with the constraint equation (\ref{const3}). 
Note that the zero curvature equation Eq. (\ref{uvpair}) also agrees with 
Eq. (\ref{sitzero}). If we parametrize the $ SU(2)$ matrix $g$ by
\be
g=e^{i\eta \s_{3}}e^{i\varphi (\cos{\q }\s_{1} -\sin{\q }\s_{2})}e^{i\eta 
\s_{3}}= \pmatrix{ e^{2i\eta }\cos{\varphi } & i\sin{\varphi }e^{i\q } 
\cr i\sin{\varphi }e^{-i\q } & e^{-2i\eta }\cos{\varphi } }\ ,
\label{para}
\ee
we recover the parametrizations of $E , P $ and $D$ as given in 
Eq. (\ref{csgepd}) and the Maxwell equation becomes the complex sine-Gordon 
equation in Eq. (\ref{csg}). The potential term in Eq. (\ref{action2}) now 
changes into the population inversion $D$,
\be
S_{\mbox{pot}} = \int {\b \over \pi }\cos{2\varphi } = 
\int {\b \over \pi }D   ,
\label{twopot}
\ee
which for $\b >0 $ possesses degenerate vacua at
\be
\varphi = \vf_{n} = (n+ {1\over 2} )\pi , \ n \in Z  \ \mbox{ and } \  
\q \ = \q_{0} \ \ \mbox{for} \ \ \q_{0} \  \mbox{constant} \ .
\label{twovac}
\ee
The property of degenerate vacua and the corresponding soliton solutions 
will be considered in Sec. 4.
\subsection{ Degenerate two-level system}
One of the deficiencies of the SIT model of McCall and Hahn is the 
absence of level degeneracy. Since most atomic systems possess level 
degeneracy, the analysis of the nondegenerate two-level system does not 
apply to a more practical system. Moreover, level degeneracy in general 
breaks the integrability and does not allow exact soliton 
configurations. For example, propagation of pulses in a two-level 
medium with the transition $j_{b} = 2 \rightarrow j_{a} = 2$ 
is effectively described by the double sine-Gordon equation
\be
\pp\pb \vf = c_{1} \sin {\vf } + c_{2} \sin{2 \vf }
\label{dsg}
\ee
which is not integrable. Nevertheless, there are a few exceptional 
cases which are completely integrable even in the presense of level 
degeneracy. It was shown that \cite{maim,bash1} the Maxwell-Bloch 
equations for the transitions 
$j_{b} = 0 \rightarrow j_{a} = 1 , \ j_{b} = 1 \rightarrow j_{a} = 0 $ 
and $j_{b} =1 \rightarrow j_{a} = 1 $ (see Fig. 1b and 1c) are 
integrable in the sense that they can be expressed in terms of 
$U-V$ pairs. In the following, we show that these cases 
correspond to the effective theory with $G = SU(3) $ and $ H = U(2) 
\subset G $. Also, we show that the local vector gauge structure 
incorporates naturally the effects of frequency detuning and 
longitudinally applied magnetic field. Consider a monochromatic pulse 
propagating through a medium of degenerate two-level atoms in the 
presence of a longitudinal magnetic field. Then, the Maxwell-Bloch 
equation under SVEA is given by 
\begin{eqnarray*}
\pb \ve ^{q} = i \sum_{\m m} \langle R_{\m m} \rangle J^{q}_{\m m} 
\end{eqnarray*}
\begin{eqnarray*}
[\pp + i ( 2 \xi  + \O_ {b} \m - \O_ {a} m )] R_{\m m } =
 i \sum_{q} \ve ^{q}( \sum_{m^{'}} J^{q}_{\m m^{'}} R_{m^{'} m} - 
 \sum_{\m ^{'}} R_{\m \m ^{'}} J^{q}_{\m ^{'} m}   )
\end{eqnarray*}
\begin{eqnarray*}
[\pp + i\O_ {a}(m - m^{'}) ] R_{m m^{'}} =
i \sum_{q \m } (\ve ^{q *} J^{q}_{\m m } R_{\m  m^{'} }  - 
 \ve ^{q } J^{q}_{\m m^{'} } R_{ m \m  } )
\end{eqnarray*}
\be
[\pp + i\O_ {b}(\m - \m^{'})] R_{\m \m^{'}} = 
i \sum_{q m } (\ve ^{q } J^{q}_{\m m } R_{m  \m^{'} }  - 
 \ve ^{q *} J^{q}_{\m^{'} m } R_{ \m m  } ) .
\label{mb1}
\ee
The dimensionless quantities ${\bf \ve }^q $ and $ { R }$ are 
propotional to the electric field amplitude ${ E}$ 
and the density matrix ${\bf \r }$, 
where $q$ is the polarization index and the subscripts $\m , 
\m^{'} , \dots $ and $ m , m^{'} , \dots $ denotes projections of the 
angular momentum on the quantization axis in two-level states $ |a \rangle $ 
and $ |b \rangle $ respectively.\footnote{For details of propotionality 
constants and their physical meanings, we refer the reader to Ref.  
\cite{maim2}.}  $J$ denotes the Wigner's $3j$ symbols
\be
J^{q}_{\m m} = (-1)^{j_{b}-m} \sqrt{3} \pmatrix{ j_{a} & 1 & j_{b} \cr
-m & q & \m },
\label{wigner}
\ee
and $\O _{a} (\O _{b} )$ is a dimensionless coupling constant of an 
external magnetic field.
In general, Eq. (\ref{mb1}) is not integrable. However, with particular 
choices $j_{a}$ and $j_{b}$, Eq. (\ref{mb1}) can be recasted into the zero 
curvature form, or the $U-V$ pair as in Eq. (\ref{uvpair}).
Specifically, for the transition 
$j_{b} = 1/2 \rightarrow j_{a} = 1/2$ (Fig. 1d), we have
\be
U = \pmatrix{ U_{+} & 0 \cr 0 & U_{-} } ~ , ~ V = \pmatrix{ V_{+} & 0 
\cr 0 & V_{-} } 
\ee
where
\ben
U_{\pm } &=& \pmatrix{ -i(x + \l ) & \pm i\ve ^{\pm 1} \cr \mp i 
\ve^{\pm 1 * } & i(x + \l ) } 
~ , ~ V_{\pm } = -{1\over 2\l }\pmatrix{R^{(b)}_{\mp{1\over 2} 
\mp{1\over 2} } &  R^{(ba)}_{\mp{1\over 2} \pm{1\over 2} } 
\cr R^{(ba)*}_{\mp{1\over 2} \pm{1\over 2} } & 
R^{(a)}_{\pm{1\over 2} \pm{1\over 2} } } \nonumber \\
x &=& {1\over 4}(\O _{a} + \O _{b} - 4\xi  ) .
\label{uvhalf}
\een
In the context of field theory, we identify the $U-V$ pair in terms of 
$g$ by
\be
U = - \gi \pp g - \gi A g - \b\l T, ~~~~ V = - {1 \over \l }\gi \Tb g 
\ee
where the gauge choice is
\be
A= \pmatrix{-ix & 0 & 0 & 0 \cr
          0 & ix & 0 & 0 \cr
         0 & 0 & -ix & 0 \cr
         0 & 0 & 0 & ix }  , ~~ \Ab = 0,
\ee
and
\be
T = -{\bar T} = {i} \pmatrix{ \s_3 & 0 \cr 0 & \s_3 }
\ee
with the Pauli matrix $\s_3$. Here, we set $\b =1$ for convenience.
The resulting field theory is specified 
by  the coset $G/H = (SU(2) \times SU(2))/ (U(1) \times U(1))$
such that $g = \pmatrix{ g_1 & 0 \cr 0 & g_2 }$ with $g_1, g_2 
\subset SU(2)$ and the two $U(1)$ subgroups are generated by 
$\pmatrix{ \s_{3} & 0 \cr 0 & 0 } $ and $\pmatrix{ 0 & 0 \cr 0 & 
\s_{3} }$.  Note that the specific form
of the identification in Eq. (\ref{uvhalf}) requires $g_1$ and $g_2$ to be
$SU(2)$ matrices as in the case of the nondegenerate two-level system. Thus, 
this case is identical to two sets of the nondegenerate two-level 
system.
\\

Another integrable case is for the transitions; $j_{b} = 1 \rightarrow 
j_{a} = 0 $ or $ j_{b} = 0 \rightarrow j_{a} = 1 $.
In each case, the $U-V$ pair is given by
\ben
U &=& \pmatrix{ -{4\over 3}i\l + i(x+y)& -i\ve^{-1} & -i\ve^{1} \cr
         -i\ve^{-1*} & {2\over 3}i\l -ix & 0 \cr
         -i\ve^{1*} & 0 & {2\over 3}i\l -iy }
\nonumber \\
&& \nonumber \\
x&=& -\O_{a} - {2\over 3}\xi  , ~~ y = \O_{a} - {2\over 3}\xi  ~ ~ ~ 
\mbox{for} ~ j_{b} = 0 \rightarrow j_{a} = 1 \nonumber \\
x&=& -\O_{b} - {2\over 3}\xi  , ~~ y = \O_{b} - {2\over 3}\xi  ~ ~ ~ 
\mbox{for} ~ j_{b} = 1 \rightarrow j_{a} = 0 ,
\label{Uonly}
\een
and
\ben
V &=& {i \over 2\l }\pmatrix{ R^{(b)}_{00} & R^{(ba)}_{0-1} & 
R^{(ba)}_{01} \cr R^{(ba)*}_{0-1}& R^{(a)}_{-1-1} & 
R^{(a)}_{-11} \cr R^{(ba)*}_{01} & R^{(a)}_{1-1} & R^{(a)}_{11}}
~ \mbox{ for } ~ j_{b} = 0 \rightarrow j_{a} = 1 \nonumber \\
&& \nonumber \\
V&=& {i \over 2\l }\pmatrix{ -R^{(a)}_{00} & R^{(ba)}_{10} & 
R^{(ba)}_{-10} \cr R^{(ba)*}_{10}& -R^{(b)}_{11} & -R^{(b)}_{-11} 
\cr R^{(ba)*}_{-10} & -R^{(b)}_{1-1} & -R^{(b)}_{-1-1}}
~ \mbox{ for } ~ j_{b} = 1 \rightarrow j_{a} = 0.  
\label{Vonly}
\een
The gauge choice is given by
\be
A= \pmatrix{-i(x+y)  & 0 & 0 \cr
          0 & ix  & 0 \cr
         0 & 0 & iy }, ~~ \Ab = 0 .
\ee
Thus, the field theory is specified by $ G/H = SU(3)/U(2)$ with
\be
T = -{\bar T} = {2 i \over 3} \pmatrix {2 & 0 & 0 \cr 0 & -1 & 0 \cr
0 & 0 & -1 }
\ee
and $\b =1$. It is interesting to observe that the electric field 
components $\ve^{1} , \ve^{-1}$ in Eq. (\ref{Uonly}) parametrize the coset 
$SU(3)/U(2)$ and the vector $(\ve^{-1*} , \ve^{1*})^{T}$ transforms as 
a vector under the $U(2)$ action. In particular, since frequency 
detuning amounts to the global $U(1) ( \subset U(2)) $ action while 
longitudinal magnetic field amounts to the global $U(1) \times U(1) 
(\subset U(2))$ action, the effects of both detuning and magnetic 
field to $\ve^{1} , \ve^{-1}$ can be easily obtained.
\subsection{ Three level system}
The propagation of pulses in a multi-level medium
with several carrier frequencies as given in Eq. (\ref{elec})   
is a more complex problem than the two-level case and in general the 
system is not exactly integrable. However, with certain restrictions on the 
parameters of the medium, it becomes integrable again and reveals 
much richer structures. Typical integrable three-level systems are 
either of $\L$-type or $V$-type as in Fig. 1e and Fig. 1f.  The $U-V$ pair for each 
system is essentially the same as that of the degenerate two-level 
system in Eq. (\ref{Uonly}). Instead of giving an explicit $U-V$ pair 
using a density matrix, we present an equivalent expression of the 
Maxwell-Bloch equation for the $\L$ or $V$-system and the $U-V$ pair 
in terms of probability components. It is given by the  the 
Schr\"{o}dinger equation
\ben
\pp c_{1}  &=& i \O_{1} c_{3} 
\nonumber \\
\pp c_{2}  &=& i \O_{2} c_{3} 
\nonumber \\ 
\pp c_{3}  &=& i( \O_{1}^{*} c_{1} + \O_{2}^{*}c_{2} ),
\label{Sch} 
\een
and the Maxwell equation
\ben
i \pb \O_{1} &=& 
s_{1} c_{1}c_{3}^{*} \nonumber \\
i \pb  \O_{2} &=& 
s_{2} c_{2}c_{3}^{*} ,
\een
where $s_{i} =  2\pi N \m_{i}^{2} \o_{i} / \hbar; ~ i = 1,2 $ and 
$c_{k};~ k=1,2,3$ are slowly varying probability amplitudes for the 
level occupations, $\O_{i} = \m_{i} E_{i}/2 \hbar $ are the Rabi 
frequencies for the transitions $i \rightarrow 3$. 
$E_{1}$ and $E_{2}$ are the slowly varying electromagnetic field 
amplitudes,  $\m_{i}$ is the dipole matrix element for the relevant 
transition and $\o_{i}$ is the corresponding laser frequency, 
and $N$ is the density of resonant three-level atoms.If the oscillator 
strengths are equal $(s_{1} = s_{2}=s)$, these equations 
can be put in the $SU(3)/U(2)$-context with the following \
identifications;
\be
g = \pmatrix{ *& * & * \cr c_{1}^{*} & c_{2}^{*} & c_{3}^{*}  \cr 
 * & * & * }     
\ee  
and
\be
g^{-1} \pp  g =  \pmatrix{ 0 & 0 & -i \O_{1} \cr      
0 & 0 & -i \O_{2} \cr - i\O_{1}^{*} & -i\O_{2}^{*} & 0}  .
\ee 
The gauge choice is that $A=\Ab=0$ and $T=\mbox{diag}(-i/2, -i/2, i/2)$.         
The density matrix $\rho$, with components $\rho_{mk} = c_{m}c_{k}^{*}$, 
is given by 
\be
\rho = -i  g^{-1} \Tb g, ~~~ 
\Tb = \pmatrix{0 &0 &0 \cr 0 & i  & 0 \cr 0 & 0 & 0} .
\label{density}
\ee 
Finally, the $U-V$ pair is given by
\be
U=-g^{-1} \pp  g - s \l T, ~~ V = - {1 \over \l} g^{-1} \Tb g .
\ee
This system of integrable equations exhibit many interesting exact 
solutions. Detailed studies of this case will appear in a separate 
paper \cite{ps3}.  Recently, three-level $\L$ and $V$-systems have 
received much attention in the context of quantum coherence effects, 
such as lasing without inversion and electromagnetically induced 
transparency. In particular, there have been extensive studies, both 
analytical and numerical, on the propagation of matched pulses through 
absorbing media \cite{harr}-\cite{Vem2}. Though our matrix potential 
formulation applies only to the nonabsorbing medium case, the exact 
analytic solutions could provide a guideline for numerical studies 
in absorbing, non-integrable cases. It is important to note that the group 
symmetry persists even in the absorbing case which leads to 
interesting results \cite{Vem2,ps3}.
\\

The degenerate three-level case and its integrability has been 
studied earlier in the context of the inverse scattering 
method \cite{Bash2}. We suppress the general Maxwell-Bloch 
equation formulation for the three-level case and refer the reader to 
Ref. \cite{Bash2} for details. Here, we extend the Maxwell-Bloch 
equation of Ref. \cite{Bash2}  to include a longitudinal 
magnetic field. Then, the Maxwell-Bloch equation describing the 
$\L $ configuration with $j_{b} = 1, j_{a} = 
j_{c} = 0 $ (Fig. 1g) is given  in a dimensionless form by
\be
\pb \ve^{q}_{j} = -i p^{q}_{j}, ~~~ j = 1,2, ~~~ q = \pm 1
\ee
and
\ben
( \pp + i t_{0}(k_{1}v - 2 \D _{1} - \O _{b} q)) p^{q}_{1} &=& 
-i(\sum_{q^{'}} \ve_{1}^{q^{'}}m_{q^{'}q} - \ve_{1}^{q} n_{1} - 
\ve^{q}_{2}r ) \nonumber \\ ( \pp + i t_{0}(k_{2}v - 2 \D _{2} - 
\O _{b} q) ) p^{q}_{2} &=& -i(\sum_{q^{'}} \ve_{2}^{q^{'}}m_{q^{'}q} - 
\ve_{2}^{q} n_{2} - \ve^{q}_{1}r^{*} ) \nonumber \\
( \pp - i t_{0}(k_{2}v -k_{1}v - 2 \D _{2} +2 \D_{1} ) ) r &=& 
-i\sum_{q} (\ve_{1}^{q}p_{2}^{q*} - \ve_{2}^{q*} p_{1}^{q} ) 
\nonumber \\
\pp n_{j} &=& -i \sum_{q} ( \ve^{q}_{j}p^{q*}_{j} - 
\ve^{q*}_{j}p^{q}_{j})  \nonumber \\
( \pp + it_{0} \O_{b} (q-q^{'} ) ) m_{qq^{'}} & = & -i 
\sum_{j=1,2}( \ve^{q*}_{j}p^{q^{'}}_{j} - \ve^{q^{'}}_{j}p^{q*}_{j}) 
\label{threemb}
\een
where $\ve_{j}^{q}, ~~j= 1,2$ is the amplitude of a double-frequency 
ultrashort pulse and $q = \pm 1$ denote the right(left)-handed 
polarization. Other variables are proportional to the components of 
the density matrix 
\ben
p_{1}^{q} &=& \rho ^{(ba)}_{-q0}\exp [-i(k_{1}x-w_{1}t)]/N_{a},~~
p_{2}^{q} = \rho ^{(bc)}_{-q0}\exp [-i(k_{2}x-w_{2}t)]/N_{a} 
\nonumber \\
n_{1} &=& -\rho _{00}^{(a)}/N_{a}, ~~ 
n_{2} = -\rho _{00}^{(c)}/N_{a} , ~~
m_{qq^{'}} = -\rho^{(b)}_{-q^{'}-q}/N_{a}  \nonumber \\
r &=& - \rho ^{(ca)}_{00}\exp [i (k_{1}-k_{2})x-i(w_{1}-w_{2})t ]/N_{a} 
\een
and $t_{0}$ is a constant with the dimension of time and $N_{a}$ is the 
polulation density of the level $|a \rangle $. 
$2 \D_{1} \equiv w_{1} - w_{ba}, ~ ~ 2 \D_{2} \equiv w_{2} - w_{bc} $ 
measure the amount of detuning from the resonance frequencies. 
The integrability of Eq. (\ref{threemb}) comes from its equivalent zero 
curvature form with the $4 \times 4$ matrix $U-V$ pair,
\ben
U &=& \pmatrix{-A_{1} - i\l {\bf 1}_{2 \times 2} & -i E \cr 
          -iE^{\dagger } & -A_{2} + i\l {\bf 1}_{2 \times 2} }
\nonumber \\
&& \nonumber \\
V &=& {i \over 2\l }\pmatrix{ -M & P \cr P^{\dagger } & -N }
\een
where 
\begin{eqnarray*}
E &=& \pmatrix{\ve^{-1}_{1} & \ve^{-1}_{2} \cr \ve^{1}_{1} 
& \ve^{1}_{2} }, ~ 
P = \pmatrix{ p^{-1}_{1} & p^{-1}_{2} \cr p^{1}_{1} & p^{1}_{2} } , ~ 
M = \pmatrix{m_{-1-1} & m_{1-1} \cr m_{-11} & m_{11} } \nonumber \\
&& \nonumber \\
N &=& \pmatrix{n_{1} & r^{*} \cr r & n_{2} }, ~ A_{1} = \pmatrix{a & 0 
\cr 0 & b}, ~ A_{2} = \pmatrix{ x & 0 \cr 0 & y }
\nonumber
\end{eqnarray*}
with
\ben
a &=& {i t_{0} \over 4}(k_{1}v + k_{2}v - 2 \D_{1} - 2 \D_{2} + 4\O_{b} )
\nonumber \\
b &=& {i t_{0} \over 4}(k_{1}v + k_{2}v - 2 \D_{1} - 2 \D_{2} - 4\O_{b} ) 
\nonumber \\
x &=& {i t_{0} \over 4}( -3 k_{1}v + k_{2}v + 6 \D_{1} - 2 \D_{2} ) 
\nonumber \\
y &=& {i t_{0} \over 4}( k_{1}v -3 k_{2}v - 2 \D_{1} +6 \D_{2} ) .
\label{degthe}
\een
Thus, in our field theory context, this corresponds to the case 
where $G/H = SU(4) /S(U(2) \times U(2))$ and the gauge choice,
\be
A = \pmatrix{ A_{1} & 0 \cr 0 & A_{2} }, ~ \Ab = 0 ,
\ee 
where $A_{1}, A_{2}$ are as in Eq. (\ref{degthe}) 
and
\be
T = -\bar{T} = {i} \pmatrix{ {\bf 1}_{2 \times 2} & 0 \cr 0 & 
-{\bf 1}_{2 \times 2} }.
\ee
Similarly, we may repeat an identification for the case \cite{Bash3}, 
$j_{a} = j_{c} = 1, j_{b} = 0$, (Fig. 1h) and can easily verify that it 
corresponds to the symmetric space $SU(5)/U(4)$. 

\section{Solitary pulses }
\setcounter{equation}{0}
The lossless propagation of optical pulses in multi-level atomic media has 
been a subject of intensive study since the discovery of self-induced 
transparency. Most theoretical works on this subject have resorted to the 
method of inverse scattering. The $2\pi $-pulse of self-induced transparency 
and its generalizations to multi-level cases, e.g. simultons, are identified 
with ``solitons" in the context of inverse scattering. Though the inverse 
scattering method is powerful enough to generate exact solutions and predict 
the evolution of a pulse of arbitrary shape, it does not explain the 
topological nature of solitary pulses. In the sine-Gordon limit, the 
$2\pi $-pulse has been identified with the topological soliton of the 
sine-Gordon theory, which is stable due to the topological number conservation. 
The topological number is protected since its change costs infinite energy. 
The cosine potential energy in the sine-Gordon theory indeed measures the 
atomic energy through the population inversion. However, except for the 
sine-Gordon limit, such a topological treatment of optical pulses was not 
possible since field theories for more general cases were absent. Therefore, 
our field theory formulation allows a topological treatment of multi-level 
optical pulses. In the following, we show in detail that the potential 
energy term in Eq. (\ref{potent}) possesses infinitely many degenerate vacua
and leads to topological solitons. In certain cases, a topological soliton is 
characterized by more than one topological number, which is a new feature 
of multi-level pulses. On the other hand, we show that there exist also 
nontopological pulses which otherwise possess all the properties of solitons. 
A new, nontopological charge is introduced for such pulses from the 
``global axial $U(1)$-gauge symmetry" of the field theory action in 
Eq. (\ref{action2}). Explicit nontopological soltions are constructed and 
identified with self-detuned solitary pulses. The nontopological 
charge measures the amount of self-detuning and the charge conservation law 
proves the stability of a nontopological soliton against small fluctuations.  
\subsection{Potential energy and topological solitons}
The potential energy term in Eq. (\ref{potent}) reveals a rich 
structure of the vacuum of the theory. It is a ``periodic" function  
in local variables. This periodicity gives rise to infinitely many 
degenerate vacua which are specified by a set of integer numbers.
Thus, any finite energy solution should interpolate between two vacua. 
In the nondegenerate two-level case, the potential term in Eq. (\ref{potential}) 
becomes a periodic cosine potential in Eq. (\ref{twopot}) and each degenerate 
vacuum is labeled by an integer $n$ as in Eq. (\ref{twovac}). A soliton
interpolating between two different vacua, labeled by $n_{a}$ and 
$n_{b}$, as $x$ varies from $-\infty $ to $\infty $  is characterized 
by a soliton number $\D n = n_{b} - n_{a}$. 
In order to understand the vacuum structure of the potential for other 
multi-level cases, we first note that the potential term $\mbox{Tr} 
(g T g^{-1} \Tb )$, charaterized by a coset $G/H$, is invariant under the 
change $g \rightarrow gh$ for $h \in H$. Consequently, we may express 
the potential term through a coset element $m \in G/H$ by
$\mbox{Tr}( m T m^{-1} \Tb ) $, where  
\be
m  = \exp  \pmatrix{ 0 &  B  \cr & \cr  -B^\dagger  &  0} =
\pmatrix{ \cos \sqrt{B B^\dagger} &  
B\sqrt{B^\dagger B}^{-1} \sin \sqrt{B^\dagger B}  \cr & \cr
 -\sin \sqrt{B^\dagger B}  \sqrt{B^\dagger B}^{-1 } B^\dagger  & 
\cos \sqrt{B^\dagger B} } .
\label{mmat}
\ee 
The matrix $B$ parametrizes the tangent space of $G/H$. This manifests 
the periodicity of the potential through the cosine and the sine 
functions. For the specific cosets, 
$ SU(2)/U(1), ~ SU(3)/U(2) $ and $ SU(4)/S(U(2) \times U(2)) $, the  
relevant matrices $B$ are complex-valued matrices of size 
$1 \times 1, ~ 1 \times 2$ and $ 2 \times 2 $ respectively. 
Owing to the relation,
\be
B \sin \sqrt{B^\dagger B} 
\sqrt{B^\dagger B}^{-1} = \sin \sqrt{B B^\dagger}  
\sqrt{B B^\dagger}^{-1} B, 
\ee
the potential term reduces to 
\be
\mbox{Tr} \left(I - 2 \sin^2 \sqrt{B B^\dagger} \right) 
+ \mbox{Tr} \left(I - 2 \sin^2 \sqrt{B^\dagger B} \right) 
\label{pot1}
\ee
for the $SU(2)/U(1)$ and the $SU(4)/S(U(2) \times U(2))$ cases and
\be
\mbox{Tr} \left(4 I - 6 \sin^2 \sqrt{B B^\dagger} \right) 
+ \mbox{Tr} \left(I - 3 \sin^2 \sqrt{B^\dagger B} \right) 
\label{pot2}
\ee
for the $SU(3)/U(2)$ case.
For a further reduction, we denote the non-zero eigenvalues of 
${B^\dagger B}$ by  $\f_i ^2 ~( i=1,.., r \equiv \mbox{rank} 
\{B^\dagger B\})$ which are positive definite and coincide with those 
of ${B B^\dagger}$. In terms of  $\f_i ^2 $, the potential term takes 
a particularly simple form, 
\be
a - b \sum _{i} \sin ^2 \f_i ,
\label{repot}
\ee
where the positive constants $a$ and $b$ can be read directly from 
Eqs. (\ref{pot1}) and (\ref{pot2}). 
In order to check, we take for example the $SU(3)/U(2)$ case and 
choose the $B$ matrix by
\be
B = (-\f \sin \h e^{-i \b} \ \ -\f \cos \h e^{-i \a})  .
\ee
Then,
\be
m = \left( \begin{array}{ccc}
\cos \f & -\sin \f \sin \h e^{-i \b} & -\sin \f \cos \h e^{-i \a} \\
\sin \f \sin \h e^{i \b} & \cos ^2 \h + \cos \f \sin^2 \h & 
-\cos \h \sin \h e^{i\b -i \a} (1 - \cos \f ) \\
\sin \f \cos \h e^{i \a} & -\cos \h \sin \h e^{ i \a - i \b} 
(1- \cos \f ) &
\sin^2 \h + \cos \f \cos^2 \h  \end{array} \right) ,
\ee
and the potential term becomes 
\be
{\rm Tr}( g T g^{-1} \Tb ) = {\rm Tr}
(m T m^{-1} \Tb )= 6 -9 \sin^2 \f 
\ee                                      
which agrees precisely with Eq. (\ref{pot2}).    
\\

The potential term in Eq. (\ref{repot}) manifests the periodicity 
of the potential and the infinite degeneracy of the vacuum.  
The minima of the potential occur at $\f_i = (n_i + 1/2) \p$  
for integer $n_{i}$. Therefore, the degenerate vacua are specified by a 
set integers $(n_{1}, n_{2}, ... ,n_{r})$.\footnote{In fact, 
in the case of multiply integer-labeled vacua, not 
all of them are topologically distinct. A similarity transformation of 
${B^\dagger B}$ which reshuffles the eigenvalues $\f_i ^2$ is a 
continuous symmetry of the vacuum, i.e. under the continuous 
similarity transformation, the potential energy does not change. 
For example, two vacua $(n_{1}, n_{2})$ and $(n_{2}, n_{1})$ 
are not topologically distinct but related by a continuous 
symmetry transformation. Also, there exists another continuous 
symmetry associated with the nontopological $U(1)$ charge which 
provides an additional topological degeneracy by the identification 
of two vacua $(n_{1}, n_{2}, ... ,n_{r})$ and 
$(- n_{1}, - n_{2}, ... ,- n_{r})$. Thus, the topological 
configuration of degenerate vacua is characterized by 
$(Z)^{r}/(Z_{r} \times Z_{2}).$} The rank $r$ of 
${B^\dagger B}$ is one for the cases of $SU(2)/U(1)$ and $SU(3)/U(2)$ 
and two for the case $SU(4)/S(U(2) \times U(2))$. Therefore, solitons for the 
$SU(4)/S(U(2) \times U(2))$ case, which interpolate between two vacua 
$(n_{1a}, n_{2a})$ and $(n_{1b}, n_{2b})$ with $|n_{1}| \ge |n_{2}| $ and 
$n_{1} \ge 0$, are labeled by two soliton 
numbers $\D n_{1} = n_{1b} - n_{1a} $ and $\D n_{2} = n_{2b} - n_{2a}$. 
In the following, we present an explicit expression for the 1-soliton 
carrying two soliton numbers. Consider the degenerate three-level system 
with the group structure $SU(4)/S(U(2) \times U(2))$. For simplicity, we 
assume that the system is on resonance ($\D_{1} = \D_{2} = 0, ~ v=0$) 
without external magnetic field and inhomogeneous broadening. This is 
equivalent to the case where $A=\Ab = 0$ in Eq. (\ref{uv}) with 
identifications in Eq. (\ref{degthe}) in terms of a $4 \times 4$ matrix $g$. 
By applying the B\"{a}cklund transformation in Ref. \cite{ps3}, 
we obtain the 1-soliton solution in terms of variables as 
in Eq. (\ref{mmat}),
\be
B = -2 B_{0}\tan ^{-1} \exp ( 2\h z + {2 \over \h }\zb + const. )
\equiv \f B_{0}
\ee
where $\h $ is a constant and $B_{0}$ is a constant $2 \times 2$ 
matrix satisfying 
\be
B_{0}B_{0}^{\dagger }B_{0} = B_{0} .
\ee
If the matrix $B_{0}$ is degenerate, i.e. $\mbox{det}B_{0} = 0$, 
it can be given in general by
\be
B_{0}={ i\over \sqrt{1 + |\alpha |^{2}}} 
\pmatrix{\theta_{1} & \theta_{2} \cr \alpha \theta_{1} & \alpha \theta_{2} }
\ee
with complex constants $\alpha $ and $ \theta_{1} , \theta_{2}$ satisfying 
$|\theta_{1}  |^{2}  + |\theta_{2}  |^{2}  = 1$. The eigenvalues of 
$B_{0}B_{0}^{\dagger } $ are then zero and one. Therefore, up to a 
global $SU(2)$ similarity  transform of $ B_{0}B_{0}^{\dagger } $, 
this solution corresponds to the (1,0)-soliton. This solution has been 
known as a simulton in earlier literatures and its scattering behavior 
has been analyzed in detail \cite{maim2}.  
For the nondegenerate $B_{0}$, we can take $B_{0}$ as an arbitrary 
$U(2)$ matrix so that $B_{0}B_{0}^{\dagger } = {\bf 1}_{2\times 2}$ 
and the corresponding solution is the (1,1)-soliton.  
This is energetically distinct from the (1,0)-soliton and also it 
can not be reached to the (1,0)-soliton via the similarity transform 
since the similarity transform preserves eigenvalues of 
$B_{0}B_{0}^{\dagger }$. Finally, physical quantities can be obtained 
from $g$ through the identification in Eq. (\ref{uv}). Explicitly,  we find 
$E$, $P$ and $M$ in Eq. (\ref{degthe}) to be
\ben
E &=& i B_{0} \pp \f = -2 i \eta B_{0} \mbox{sech} \D 
 \nonumber \\
P &=& -2 B_{0} \sin 2 \f = - 4 B_{0} \mbox{tanh} \D \mbox{sech} \D 
\nonumber \\
 M &=& -N = -2 {\bf 1}_{2 \times 2} \cos 2 \f =-2 {\bf 1}_{2 \times 2} 
 ( 1 - 2 \mbox{sech}^{2} \D ) \nonumber \\
 \D &\equiv & 2 \eta z + {2 \over \eta} {\bar z} + const. 
\een
respectively. Inclusion of detuning and external magnetic effects can 
be done easily by a gauge transform;
\be
E \rightarrow H_{1}^{-1}EH_{2}, ~ M \rightarrow H_{1}^{-1}MH_{1}, ~ 
P \rightarrow H_{1}^{-1}PH_{2}, ~ N \rightarrow H_{2}^{-1}NH_{2}
\ee
where $H_{1}, H_{2}$ are given by $A_{1}=H_{1}^{-1}\pp H_{1} , 
~ A_{2}=H_{2}^{-1}\pp H_{2}$ for $A_{1}, A_{2}$ in Eq. (\ref{degthe}).
                
\subsection{Nontopological solitons as self-detuned pulses} 
Here, we address the issue of topological vs. nontopological solitons in 
optical systems. In order to facilitate the problem, we first focus on 
the $2\pi$ pulse of the nondegenerate two-level system. 
Set $\b =1$ in Eq. (\ref{sit}) without loss of generality. Then, by using 
the dressing method in the Appendix A, one can obtain the $2\pi$ pulse 
solution such that
\ben
\cos{\vf } &=& {b \over \sqrt{(a-\x )^{2} + b^{2} }} 
\mbox{sech} (2bz -2b C_{1} \zb ) \nonumber \\
\q &=& - \tan^{-1}[ {a-\x \over b } \mbox{coth} (2bz - 2bC_{1}\zb )] - 
2\x z + (2(a-\x )C_{1} - 2C_{2})\zb  \nonumber \\ 
\h &=& (a-\x )z + (a-\x ) C_{1} \zb    ,
\label{solisol}
\een
where $a, b$ are arbitrary constants and 
\be
C_{1} = \left< { 1\over (a-\x^{'} )^{2} + b^{2}} \right> \ ,
 \ C_{2} = \left< {  a-\x^{'} \over (a-\x^{'} )^{2} + b^{2}} \right> ,
\ee
where the angular bracket is as in Eq. (\ref{inhomog}). 
In terms of $E$ as defined in Eq. (\ref{csgepd}), the $2\pi$ pulse is 
given by
\be 
E = -2ib \  \mbox{sech} (2bz - 2bC_{1} \zb ) e^{-2i(az + C_{2}\zb )} .
\label{esoliton}
\ee
Note that $E$ is explicitly independent of $\x $ in Eq. (\ref{esoliton}), 
despite the $\x$-dependence of potential variables $\vf , \h $ and 
$\q$. This exemplifies the macroscopic nature of  $E$ as discussed in 
Sec. 2.3. 
In the sharp line limit of the frequency distribution $f(\xi ^{'}) = 
\d (\xi ^{'} - \xi_{0} ) $, this solution retains the same form except 
for the change of constants $C_{1}$ and $C_{2}$,
\be
C_{1} =  { 1\over (a-\x_{0} )^{2} + b^{2}}  \ , 
\ C_{2} =  { a -\x_{0} \over (a-\x_{0} )^{2} + b^{2}} .
\ee
The solution in Eqs. (\ref{solisol}) and (\ref{esoliton}) is loosely 
identified with the 1-soliton in earlier works using the inverse 
scattering method. However, this does not necessarily 
mean that it is a topological 1-soliton. We emphasize that the 
topological distinction is possible only in the sharp line limit and 
even in that case not all solitons are topological solitons. 
For example, when $a=\x$, the above solution describes a localized pulse 
configuration which interpolates between two different vacua in 
Eq. (\ref{twovac}) such that 
\be
\vf (x = -\infty) = (n + {1 \over 2}) \p , ~~
 \vf (x = \infty) = (n + {1 \over 2}-{b \over |b| }(-1)^{n} ) \p .
\ee
Thus, it carries a topological number $\D n = (-1)^{n+1}b/|b| $ and 
becomes a topological soliton. When $a \ne \x$, the solution 
reaches to the same vacuum as $x \rightarrow \pm \infty $ since the peak of 
the localized solution does not reach to the point where $\cos {\vf } = 1$. 
That is, its topological number is zero. Nevertheless, it shares 
many important properties (e.g. localization, scattering behavior etc.) 
with the topological soliton so as to deserve the name, a  
``nontopological soliton". Note that the envelope function $E$, and also 
the time area of $E$, become complex when $a \ne \x$. But the time 
area of the absolute value of $E$ in Eq. (\ref{esoliton}) is still $2\pi$.
This suggests that we could call the solution in Eq. (\ref{solisol}) as a 
$2\pi$ pulse in a broad sense, which comprises both the topological and 
the nontopological solitons as well as the inhomogeneously broadened 
solution. In order to see the physical meaning of a nontopological 
soliton, consider the resonant case where $\x = 0$. In this case, 
Eq. (\ref{esoliton}) shows that the nontopological soliton $(a \ne 0)$ 
shifts the carrier frequency by $\D w = 2a$. Thus, the nontopological 
soliton represents a self-detuned $2\pi $ pulse. It also receives a 
spatial modulation given by a phase factor 
$\mbox{exp}(-2ia\zb /\sqrt{a^2 + b^2})$. At the microscopic level, 
the maximum population inversion, $D=\cos 2 \vf$, does not reach to 1 in 
the nontopological case so that the shape of $\vf $ is not of the 
kink-type (see Fig. 2a-2d). Note that the field intensity $|E|$ in Fig. 2d is 
invariantly hyperbolic secant-type independent of the value $a$. 

\begin{figure}  
\vglue -.9in   
\leftline{\epsfysize 1.9 truein \epsfbox {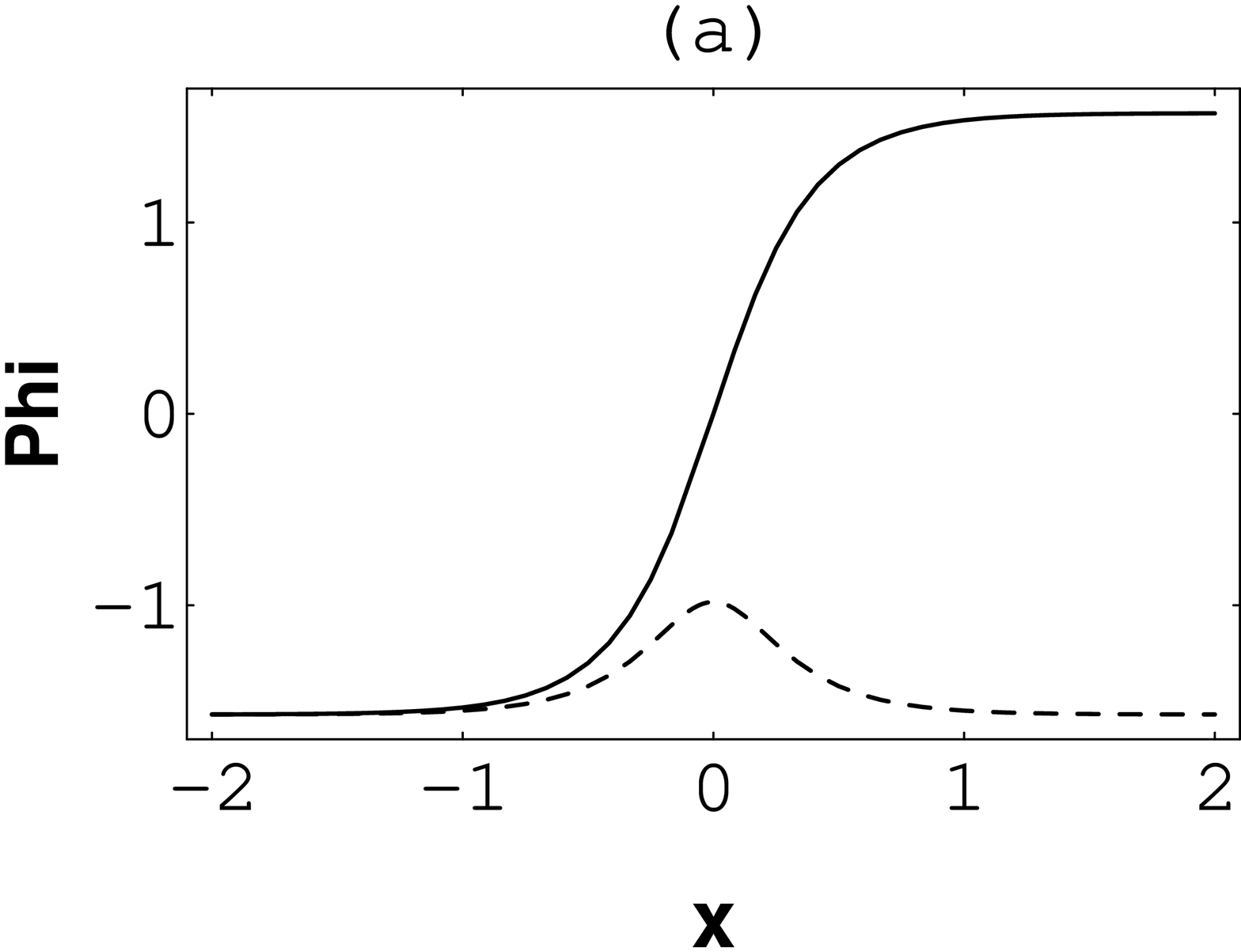}}
\vglue -1.9in
\rightline{\epsfysize 1.9 truein \epsfbox {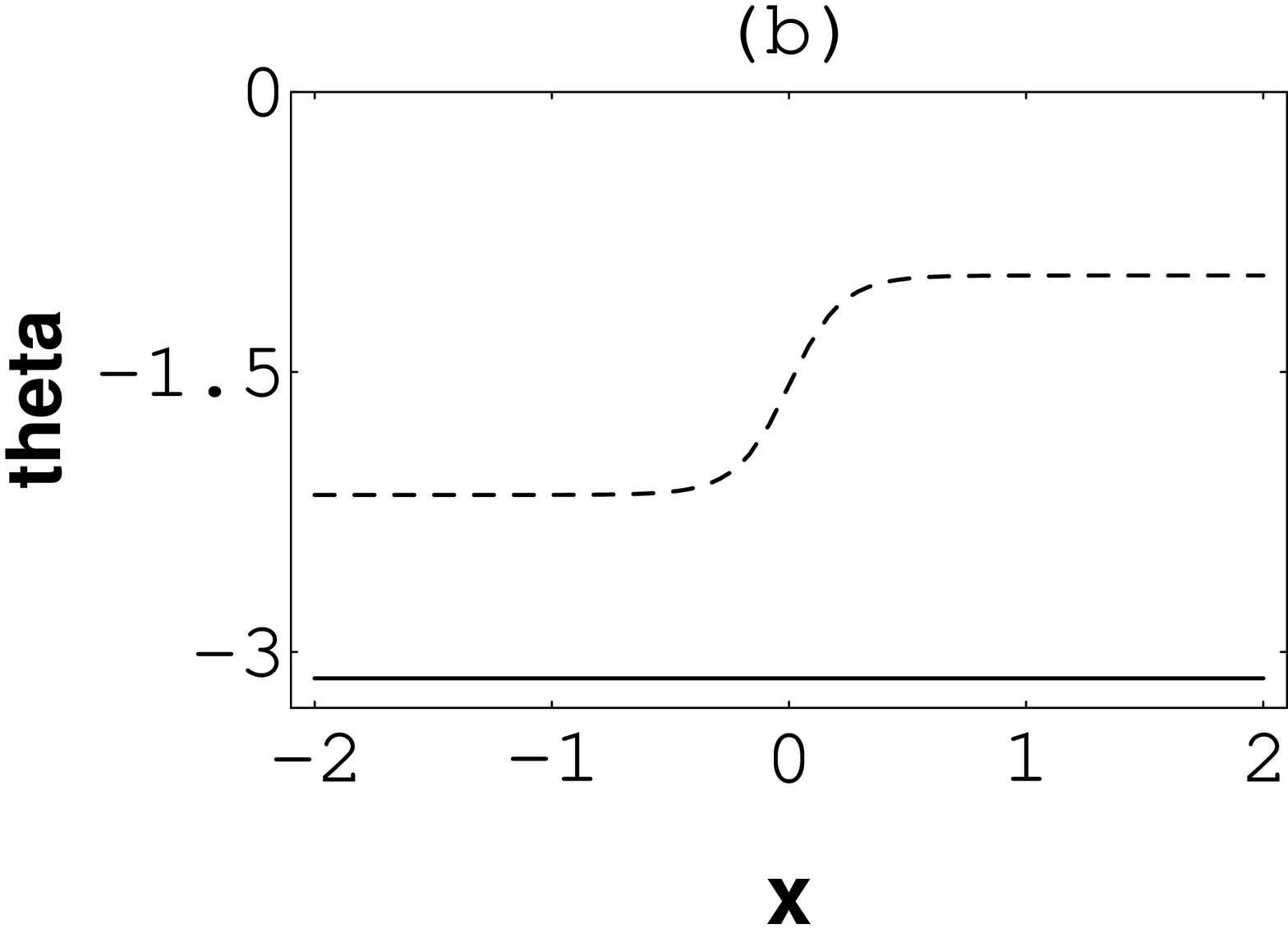}}
\vglue .3in
\leftline{\epsfysize 1.9 truein \epsfbox {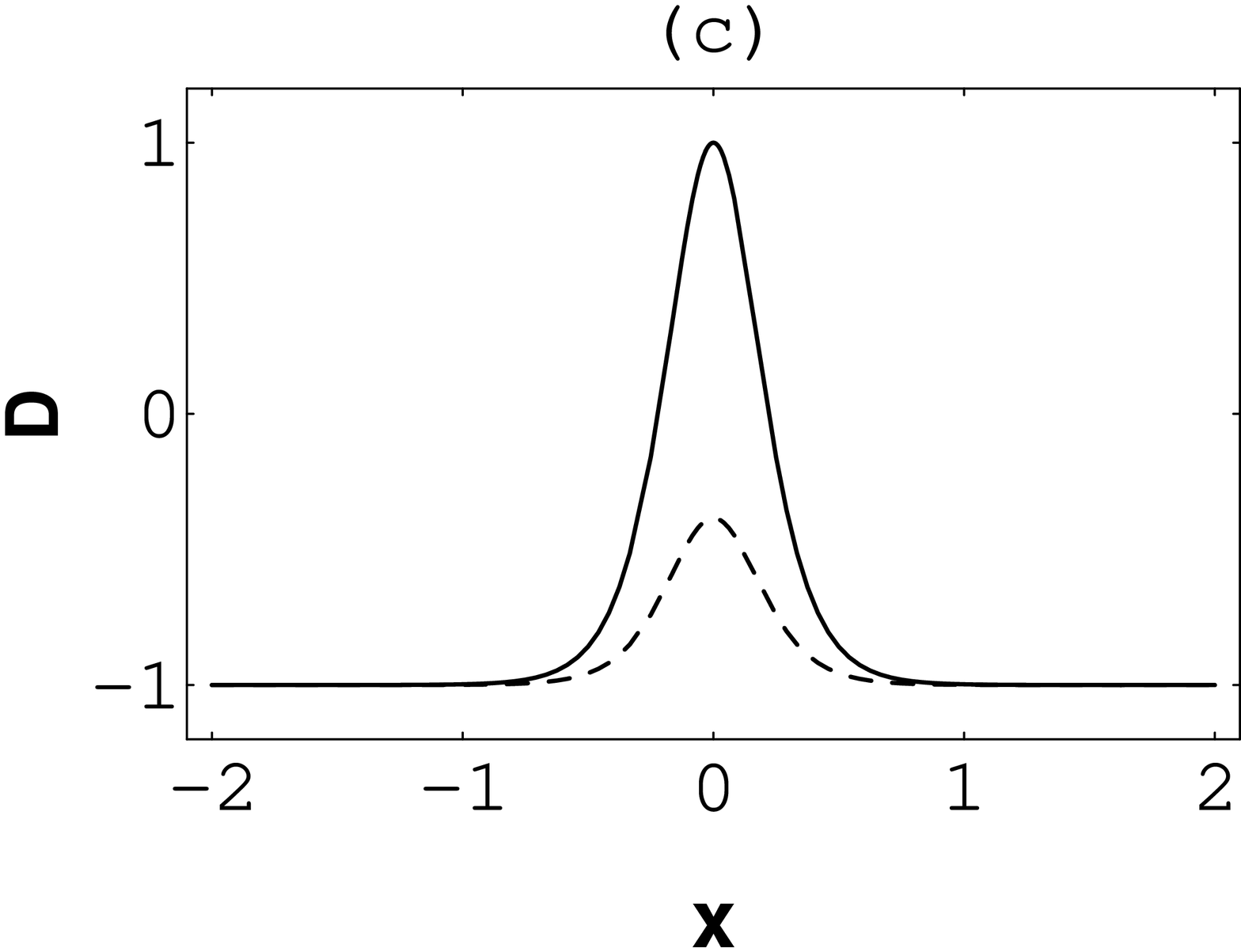}}
\vglue -1.9in
\rightline{\epsfysize 1.9 truein \epsfbox {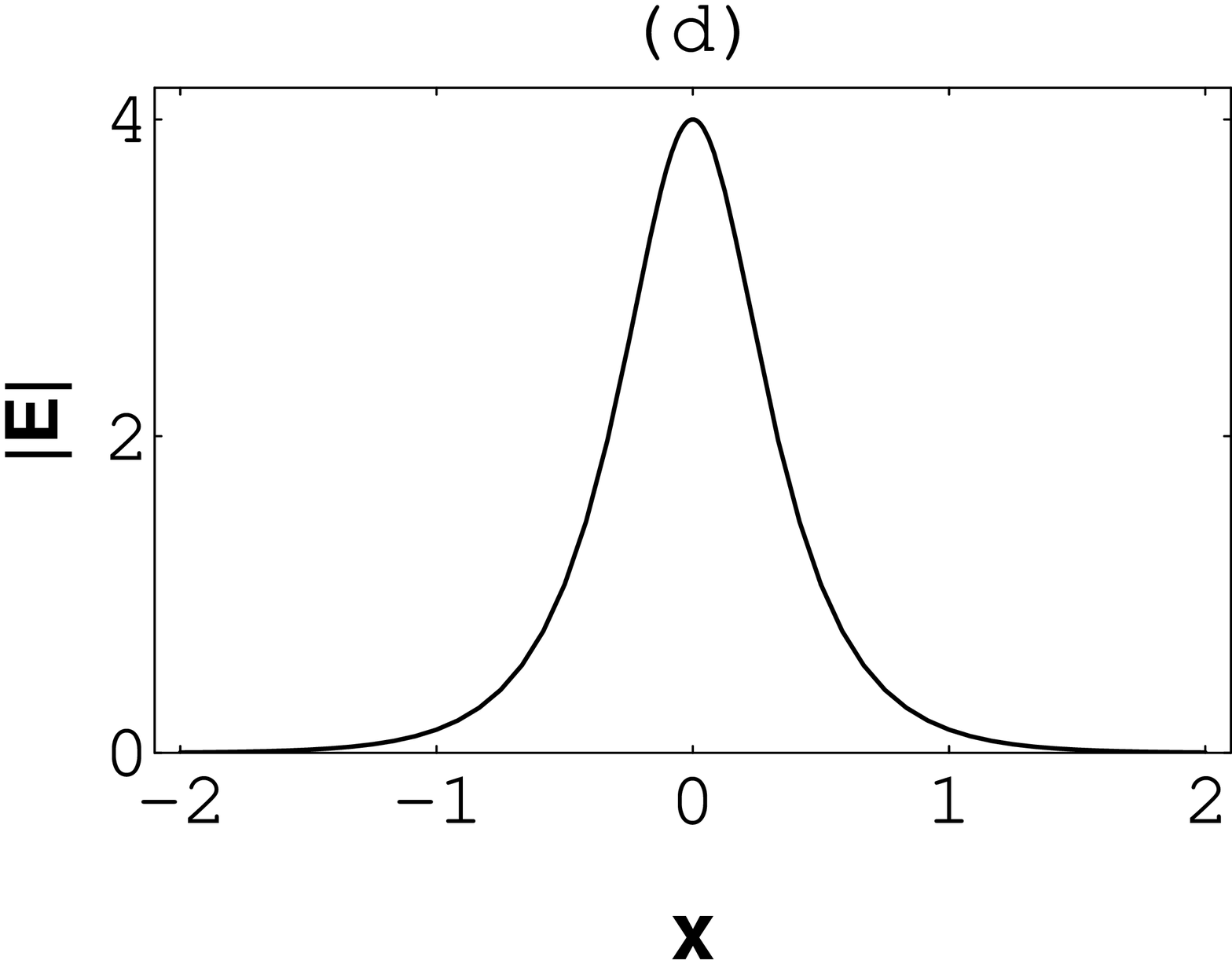}}                                             
\caption{Plots of $\f$,  $\q$ ,  $D$ and $|E|$ for (bright) one-soliton 
with $b=2 , ~ \x$=0 . 
The plot is w.r.t. $x= 2bz - 2bC_{1} \zb $. Real ($a=0$) and dashed ($a=3$) 
lines represent the topological and the nontopological solitons.
}
\end{figure}

Though a nontopological soliton can not be specified by a topological 
integer number, it carries a continuous nontopological charge. In fact, 
as we will show in the next section, the nontopological charge 
conservation law gives rise to the stability of a nontopological soliton.
In Sec. 5, we show that the symmetry leading to the nontopological charge 
is ``the global axial $U(1)$-vector gauge symmetry" of the action. In 
the nondegenerate two-level case, this means the invariance of the 
action in Eq. (\ref{csglag}) under the change 
\begin{equation}
\eta \rightarrow \eta + \g \ \   \mbox{ for } 
\g  \  \mbox{constant} .
\end{equation}
The corresponding Noether current is given by 
\be 
J = {\cos^{2}{\varphi } \over \sin^{2}{\varphi }}\partial \eta  \ \ , 
\ \ 
\bar{J} = {\cos^{2}{\varphi } \over \sin^{2}{\varphi }}\bar{\partial} 
\eta 
\label{charcur}
\ee
which satisfies the conservation law, 
\be
\partial \bar{J} + \bar{\partial} J = 
{\pp \over \pp t} [\cot^2 \vf ( \pp + \pb ) \h ] + c { \pp \over \pp x}
[\cot^2 \vf \pp \h ] = 0.  
\label{chcon}
\ee                                   
The corresponding conserved charges are either conserved in time, 
$dQ^{T}/dt=0$, 
\be
Q^{T} \equiv \int _{- \infty } ^{+ \infty } 
[ \cot ^2 \vf (\pp + \pb ) \h ] dx,
\ee
or in space, $dQ^{S}/dx=0$, 
\be
Q^{S} \equiv \int _{- \infty } ^{+ \infty } [c \cot ^2 \vf \pp \h ] dt.
\ee   
In the case of the nontopological soliton given in Eq. (\ref{solisol}), 
\be
Q^{S} = Q^{T}= c \tan^{-1}{|b| \over a- \x} .  
\label{tcharge}
\ee
The physical meaning of $Q^{S}$ is clear.  Consider a $2\pi$ pulse 
with $b$ fixed. Since $\pp \h = a - \x = a+ w_{o} -w$ expresses 
frequency detuning and $\cot ^2 \vf $ is peaked around the soliton, 
$Q^{S}$ measures precisely the amount of self-detuning of a 
nontopological soliton. Stability of nontopological solitons can 
be proved either by using conservation laws in terms of charge and 
energy as given in \cite{shin1}, or by studying the behavior against 
small fluctuations.

\subsection{Stability}
The physical relevance of a topological number is that it accounts 
for the stability of solitons against ``topological" (soliton number 
changing) fluctuations. In fact, any finite energy solution must 
approach to one of the degenerate vacua at $x = \pm \infty $ and 
therefore it carries a specific topological number. Topological 
numbers cannot change during any physical process due to the infinite 
potential energy barrier between any two finite energy solutions with 
different topological numbers. This infinite energy barrier results 
from the infinite length of the $x$-axis despite the finite potential 
energy density per unit length. On the other hand, topological number 
is not useful in understanding the stability of pulses against 
nontopological (finite energy) fluctuations. Also, the topological 
notion does not apply to the case with inhomogeneous broadening.
In Sec. 2.3, we have argued that the potential variable $g$ is microscopic 
depending on the frequency $\xi $ while the pulse amplitude $E$ is 
macroscopic being independent of $\xi $. This was apparent in the 
example of the 1-soliton solution given through 
Eqs. (\ref{solisol})-(\ref{esoliton}).
It shows that inhomogeneous broadening requires $E$ to be a function 
of ``frequency $\xi $ averaged" coefficients, while the topological vs. 
nontopological nature of the soliton critically depended on $\xi $ 
as in Eq. (\ref{solisol}). Thus, inhomogeneously broadened pulses do not 
carry topological numbers. In this regard, it is remarkable that 
the McCall and Hahn's area theorem provides a stability statement even 
in the presence of inhomogeneous broadening. In fact, the proof of the 
area theorem relied crucially on the averaging over the frequency 
$\xi$ of detuning in inhomogeneous broadening.   
However, one serious drawback of the area theorem is that it applies 
only to the case of real $E$ which ignores frequency modulation, 
and it also assumes the symmetric frequency distribution. Presently, 
a more general area theorem including frequency modulation is not known.
\\

In this section, we attempt to generalize the area theorem 
to include frequency modulation. Though we do not have the general 
area theorem, we show that how pulse reshaping with frequency 
modulation can be explained to a certain extent. In the nondegenerate 
two-level case without inhomogeneous broadening, we prove the 
stability of a $2\pi $ pulse in terms of a ``modified area function" 
and show that the recovery of soliton shape is slower in the off 
resonant case ($a-\xi \ne 0$) than in the resonant case. When 
frequency modulation is taken into account, a numerical work testing 
the pulse stability has shown that there exists a frequency-pulling 
effect \cite{diels}. This effect is explained nicely in terms of 
the nontopological charge and its conservation law.
Consider the 1-soliton in Eq. (\ref{solisol}). Since the time 
area of complex $E$ is not meaningful, instead we regard   
$\vf $ of the complex sine-Gordon equation  as a ``modified
area function". We also assume without loss of generality that 
the asymptotic time behavior of $\vf $ is given by
\ben
\vf(t= -\infty,x) =-\p /2 , ~~ \vf(t= \infty,x) =\p /2 ~ \mbox{ for } 
~ a= \x \nonumber \\
\vf(t = -\infty,x) = \vf(t = \infty,x) = -\p /2 ,~ \mbox{ for }~ a \ne 
\x .
\een
Then, the modified area 
\be
A= \int^{\infty }_{-\infty } 2 \pp \vf dt 
\ee  
of the topological soliton ($a-\x =0$) is $2\pi $ while that of the 
nontopological soliton is zero. Consider a pulse perturbed 
around the 1-soliton with the boundary condition 
$\vf (t= -\infty,x) =-\p /2$, i.e. the it is initially 
in the vacuum state. Near the trailing edge of the pulse ($t >> 1 $), 
the modified area function is perturbed by $\d \vf = \e $ for small 
$\e $, i.e. $\vf (t>>1) = \pm \pi /2 + \e$. Then, the perturbed complex 
sine-Gordon equation for $t >>1$ around the 1-soliton becomes
\be
\pb \pp \vf + 4 {b^2 \over (a-\x)^2+b^2} \e = 0 .
\label{recovery}
\ee
The perturbation of the $\eta $-part is neglected since its contribution 
is of the order $\e^{2}$.  
This shows that if the modified area is greater than $2\pi $ 
(or zero) by the amount $\e > 0$, then $\pb \pp \vf < 0$. 
Therefore, the field $\pp \vf $ at the trailing edge tends to 
decrease along the $\zb = x/c$ axis so as to recover the total 
modified area $2\pi $ (or zero). 
On the other hand, if  $\e <0 $, then 
$\pb \pp \vf > 0$ and the field at the trailing edge increases. 
This shows that the total modified area tends to remain $2\pi $ 
or zero. Moreover, Eq. (\ref{recovery}) shows that the recovery of area 
is faster in the resonant case ($a= \x$ ) than in the off-resonant 
case ($a \ne \x $).  This agrees with the prediction made by a 
numerical work \cite{diels}. In fact, the recovery of the modified 
area can be acompanied by a stronger recovery of pulse shape to that 
of a soliton. Instead of proving this, we simply point out that 
the stability of a soliton against modified area preserving 
fluctuations could be demonstrated by modifying the Lamb's 
proof in terms of the Liapunov function \cite{lamb2}, as well as 
by proving the stability of higher order conserved 
charges \cite{AKN}.  

In order to understand the frequency modulational stability, we 
recall that the nontopological charge measures the amount of 
frequency self-detuning of pulses. In the following, we show that 
the stability of the nontopological charge accounts for the 
frequency-pulling effect. 
From Eq. (\ref{chcon}), we have
\be
{dQ^{S} \over dx} = 
- \cot^2 \vf (\pp + \pb) \h |_{t= -\infty}^{t= + \infty}.
\ee
For a 1-soliton solution, the boundary contribution is zero and
$Q^{S}$ is conserved in space. If the solution is perturbed 
around the soliton such that near the trailing edge of the pulse, 
\ben
\vf (t >> 1 , x)
&=& \pm \p /2 + \e (x) \nonumber \\
\h (t >> 1 , x) &=& (a-\x )(t-{x \over c} - 
{ 1 \over (a-\x )^2 + b^2 } {x \over c}) + \d (x) 
\een 
for small parametric functions $\e (x) $ and $\d (x)$. To the leading 
order, the variation of $Q^{S}$ then becomes
\be
{d\d Q^{S} \over dx } = -(a-\x ) \left( 1 + { 1 \over (a-\x )^2 + b^2 } 
\right) \e ^2 .
\ee
This shows that the detuning by a higher frequency, i.e. $a-\x >0 $ 
reduces $Q^{S}$ for increasing $x$ while the lower frequency 
detuning does exactly the opposite. Since the conserved charge 
$Q^{S}$ of the 1-soliton is $c \tan^{-1} [ |b|/(a-\x)]$, it can be 
concluded that the absolute value of $Q^{S}(x)$ decreases monotonically 
along the $x$-axis. Eventually, it converges to a constant value of 
a soliton. Note that the monotonic decrease of $|Q^{S}|$-value of a pulse is 
slower than that of the modified area since it is of the order $\e^2$. 
The decreasing and converging behavior of $|Q^{S}|$ is in good 
agreement with the numerical work \cite{diels}, which showed that 
the frequency of the optical pulse is pulled towards the transition 
frequency and reaches to a constant value along the $x$-axis.
Thus, the stability of modified area and nontopological charge 
provides a generalization of the area theorem in the presense of 
frequency detuning in a restricted sense. A full-fledged generalization 
should include inhomogeneous broadening, in which case the 
nontopological charge conservation law breaks down. It introduces 
an anomaly term $M$ in the current conservation law, 
$\partial \bar{J} + \bar{\partial} J = M$, for $J, \bar{J}$ 
as in Eq. (\ref{charcur}) and
\begin{eqnarray} 
M &=& 2\cot{\varphi } [ \ \cos (\theta - 2\eta ) 
< \sin (\theta - 2\eta ) \sin {2\varphi } > \nonumber \\ 
&& - 
\sin (\theta - 2\eta ) < \cos (\theta - 2\eta ) \sin {2\varphi } >  
-(\cot^{2}{\varphi }\bar{\partial} \eta  + {1\over 2}\bar{\partial} 
\theta )\partial \varphi  ] .
\end{eqnarray} 
This anomaly vanishes in the sharp line limit due to the constraint 
in Eq. (\ref{constraint}). It also vanishes for the 1-soliton and the 
charge remains conserved in this case. This may be compared with the 
conserved area of topological solitons in the presence of inhomogenous 
broadening. The area theorem of McCall and Hahn proves that 
inhomogeneous broadening changes the pulse area until it reaches to 
those of $2n \pi $ pulses. This suggests that a generalized area theorem 
of pulse stability including frequency modulation may be proven by 
making use of the nontopological charge and the anomaly. But this has 
yet to be seen.

\section{Symmetries}
\setcounter{equation}{0}
One of the advantage of having a field formulation of the Maxwell-Bloch 
equation is that the field theory action reveals symmetries of 
the system. In this section, we show that our group theoretic formulation 
in particular reveals previously unknown gauge-type symmetries which have 
definite physical implications. Also, by using the group theory, we 
construct systematically infinitely many conserved local integrals of the 
Maxwell-Bloch equation in association with a Hermitian symmetric space $G/H$. 
These conservation laws can be extended to the case with inhomogeneous 
broadening without difficulty. In addition to these continuous symmetries, 
we show that the action in Eq. (\ref{action2}) uncovers two types 
of discrete symmetries; the chiral and the dual symmetries. These discrete 
symmetries relate two different solutions. In particular, we show that the 
dual symmetry relates a ``bright" soliton with a ``dark" soliton.

\subsection{Conserved local integrals}
In the Appendix B, it is shown that the associated linear 
equation (\ref{lineareqn}) in terms of a $U-V$ pair yields exact soliton 
solutions through the dressing procedure. The same linear equation can 
be employed to construct infinitely many conserved local integrals. 
We first construct conserved integrals for the $SU(2)/U(1)$ case with 
inhomogeneous broadening and later generalize to the arbitrary $G/H$ case. 
Recall that the linear equation for the $SU(2)/U(1)$ case is given by
\be
(\pp + \pmatrix{ 0 & -E \cr E^{*} & 0 } - \l T  )\Psi =0 , ~~ 
( \pb +  \left< {g^{-1}\bar{T} g \over {\lambda } + \xi 
} \right> ) \Psi = 0
\ee
where $T = - \Tb = i \s_{3} $.
We introduce the notation
\ben
 \left< g^{-1} \Tb g \right> _{l } &\equiv & 
\left< g^{-1} \Tb g (-\x )^{l}  \right> 
= -i \pmatrix{ \left< D(\xi ) (-\xi )^{l} \right> & 
\left< P(\xi ) (-\xi )^{l} \right> \cr \left< P^{*}(\xi ) 
(-\xi )^{l} \right>  &
  - \left< D(\xi ) (-\xi )^{l} \right> } \nonumber \\
&\equiv &  -i \pmatrix{ D_{l} & P_{l} \cr P_{l}^{*} & -D_{l} } ,
\een
and define 
\be
\Psi \exp(-\l T z ) \equiv \sum_{i=0}^{\infty } {1 \over \l ^{i}}\Phi_{i} 
~ ;  ~~ \Phi_{i} \equiv \pmatrix{ p_{i} & q_{i} \cr r_{i} & s_{i} }
\ee
so that the linear equation changes into
\be
(\pp + \pmatrix{ 0 & -E \cr E^{*} & 0 } ) \Phi_{i} - [T \ , \ 
\Phi_{i+1} ] = 0
\ee
and 
\be
\pb \Phi_{i} + \sum^{i-1}_{l=0} \left< g^{-1} \Tb g \right> _{i-l-1 }
\Phi_{l} = 0.
\ee
These equations can be solved iteratively in components,
\ben
q_{i} &=& {1 \over 2i} (\pp q_{i-1} - Es_{i-1} ) \\
r_{i} &=& -{1 \over 2 i} (\pp r_{i-1} + E^{*} p_{i-1} ) \\
p_{i} &=& \int Er_{i} dz + i\sum_{l=0}^{i-1}\int (D_{i-l-1}p_{l} + 
P_{i-l-1}r_{l} )d\bar{z} \\
s_{i} &=& -\int E^{*} q_{i} dz + i\sum_{l=0 }^{i-1}\int ( -D_{i-l-1} 
s_{l} +P^{*}_{i-l-1}q_{l} )d\bar{z}
\een
together with the initial conditions:
\be
p_{0} = s_{0} = -2i , \ r_{0} = q_{0} = 0.
\ee
The consistency condition, $\pp \pb p_{i} - \pb \pp p_{i} = 0$, then 
leads to the infinite current conservation laws, 
$\pp \bar{J}_{i} + \pb J_{i} =0$ for $\bar{J}_{i} = \pb p_{i}$ and 
$J_{i} = - \pp p_{i}$. Or 
\be
i\pp \sum_{l=0}^{i-1} (D_{i-l-1}p_{l} + 
P_{i-l-1}r_{l} ) - \pb ( Er_{i} ) = 0 .
\label{current}
\ee
Another consistency condition, $\pp \pb s_{i} - \pb \pp s_{i} = 0$, 
gives rise to the complex conjugate pair of Eq. (\ref{current}). 
A few explicit examples of conserved currents are
\ben
\bar{J}_{1} &=& -2D_{0} \nonumber \\ 
J_{1} &=& EE^{*} \\
\bar{J}_{2} &=& 4iD_{1} - 2P_{0}E^{*}  \nonumber \\ 
J_{2} &=& E\pp E^{*} \\
\bar{J}_{3} &=&  -2P_{0}\pp E^{*} + 8D_{2} + 4i E^{*} P_{1} \nonumber \\ 
J_{3} &=& E\pp ^{2} E^{*} + (EE^{*})^{2} \\
\bar{J}_{4} &=& -16iD_{3} + 8E^{*} P_{2} + 4iP_{1}\pp E^{*} -2P_{0} 
\pp^{2}E^{*} - 2P_{0}E^{*}|E|^{2} \nonumber \\ 
J_{4} &=& E\pp ^{3} E^{*} + |E|^{2} \pp |E|^{2} + 2E |E|^{2} \pp E^{*}  .
\een  
Half of the above integrals have appeared earlier in  \cite{lamb2}. 
As for the general $G/H$ case, we introduce the abbreviation:
\ben
\E &\equiv&  {\bar U} =g^{-1} \pp g +  g^{-1} A g 
- \x T  \nonumber \\
\left< {\bar V} \right>_l &\equiv&
\left< g^{-1} {\bar T} g \right>_l = 
\left< g^{-1} {\bar T} g (-\x)^l \right> \nonumber \\
&=& D_l +P_l,    
\label{ghabb}
\een
where in the last line, the decomposition is made according to 
the behavior under the adjoint action of $T$ such that 
$[T, ~ D_l]=0$ and $ [T, ~ P_l] \ne 0$.
Now, define matrices $\cC_{i} $ and $\cD_{i}$ recursively by
\be
\cD _{i} = - [T\ ,\ \pp \cD _{i-1}]-[T\ ,\ \E ]\cC _{i-1}
\ee
and
\be
\cC _{i} = - \int \E \cD _{i} dz 
- \sum_{l=0}^{i-1} \int (D_{i-l-1} \cC _l + P_{i-l-1} \cD _l ) d \bar z.
\ee
These matrices $ \cC _{i} $ and $ \cD _{i} $ can be 
determined completely with appropriate initial conditions. For example, 
if we choose an initial condition which is consistent with the recursion 
relation for $i \le 0$,
\be
\cC _0 = I \ \ \ , \ \ \ \cD _0 =0,
\ee
we find for the first few explicit cases in the series,
\be
\cC _1 = \int \E[T\ ,\ \E ] dz -\int D_0 d \bar z , ~~~
\cD _1 = -[T\ ,\ \E ] 
\ee
and 
\ben
\cC _2 &=& \int (\E \pp \E + \E [T\ ,\ \E ] \cC _1 ) dz
+ \int (-D_1-D_0 \cC _1 + P_0[T\ ,\ \E ]) d \bar z 
\nonumber \\ 
\cD _2 &=& - \pp \E - [T\ ,\ \E ] \cC _1  .
\een
These matrices give rise to infinitely many conserved local currents,
\be
J_{i} \equiv \pp \cC _{i} = - \E \cD _{i} 
, ~~~~ {\bar J}_{i} \equiv \pb \cC _{i} =  
- \sum_{l=0}^{i-1}  (D_{i-l-1} 
\cC _l + P_{i-l-1} \cD _l ) ,
\ee
which satisfy $\pp {\bar J}_{i} =\pb J_{i}$. An explicit derivation of 
these currents are given in the Appendix B. A few examples of currents 
are
\be
J_1 = \E [T\ ,\ \E] \ \ , \ \ {\bar J}_1 = -D_0
\ee
\be
J_2 = \E \pp \E + \E [T\ ,\ \E ] \cC _1 \ \ , \ \ 
{\bar J}_2 = -D_1-D_0 \cC _1 + P_0[T\ ,\ \E ].
\ee
The first current $J_{1} , ~{\bar J}_{1}$ gives rise to the energy 
conservation law.

\subsection{Global gauge symmetries}
The action in Eq. (\ref{action2}) for the coset $G/H $ possesses 
various type of gauge symmetries. Since the Maxwell equation arises 
from the action, it also possesses gauge symmetries, while the Bloch 
equation could change under the gauge transformation. For example, the 
local $H$-vector gauge symmetry, as given in Eq. (\ref{gaugetr}) where 
the local function $h$ belongs to the subgroup $H$, is a symmetry of 
the Maxwell equation, while the Bloch equation changes under the 
transformation. In fact, it was shown that a particular local gauge 
fixing accounts for the effect of frequency detuning and external 
magnetic fields. On the other hand, even after the local gauge fixing, 
there remains global gauge symmetries. For example, assume the local 
gauge choice $A=\Ab=0$. Then the action in Eq. (\ref{action2}) possesses 
the property
\be
S_{MB}(LgR) = S_{MB}(g)
\ee
for constant matrices $R$ and $L$ which commute with $T$ and $\Tb$ 
respectively. Thus, $R$ is an element of the subgroup $H$. 
Under the transformation $g \rightarrow LgR$, electric field components 
the density matrix components rotate among themselves via the similarity 
transform
\be 
\gi \pp g \rightarrow R^{-1} (\gi \pp g) R, ~~ 
\gi \Tb g \rightarrow R^{-1} (\gi \Tb g) R   
\ee
For example, in the case of three-level $\L $ or $V$-systems, the 
Rabi frequency $\O_{i}$ and the probability amplitude $c_{i}$ are rotated 
by
\be
\pmatrix{ \O^{'}_{1} \cr  \O^{'}_{2} } 
= \pmatrix{ h_{11}^{*} & h_{21}^{*} \cr h_{12}^{*} & h_{22}^{*} } 
\pmatrix{ \O_{1} \cr \O_{2}}, 
~~~ \pmatrix{ c^{'}_{1} \cr c^{'}_{2} \cr  c^{'}_{3} }
= \pmatrix{ h_{11}^{*} & h_{21}^{*} & 0  \cr h_{12}^{*} & h_{22}^{*} & 0 \cr 
0 & 0 & h_{33}^{*}} \pmatrix{ c_{1} \cr c_{2} \cr c_{3} } ,            
\label{sym}
\ee     
in which case 
\be
R = 
\pmatrix{ h_{11} & h_{12} & 0 \cr h_{21} & h_{22} & 0 \cr 0 & 0 & h_{33} }.
\ee
Note that when $  \O_{1} =0 $ and $ \O_{2}$ is a $2\pi$ sech pulse, the 
rotated Rabi frequencies are all propotional to the $2\pi$ sech pulse, i.e. 
it becomes a simulton solution. Thus, our global symmetry provides a 
systematic way to generate simulton solutions. 
When $L=R= \exp(\g T) \in U(1)$, we have the global $U(1)$-axial vector 
symmetry. The Noether charge of this $U(1)$-invariance is precisely the 
nontopological conserved charge introduced in Sec. 4.2. Even though a 
general expression for the nontopological charge should be possible, 
in practice it requires an explicit (noncompact) parametrization of the 
group variable $g$ as in the case of Sec. 4.4. 
\subsection{Discrete symmetries}
Besides continuous symmetries, the action in Eq. (\ref{action2}) also 
reveals discrete symmetries, {\it the chiral symmetry} and 
{\it the dual symmetry}. They are manifested most easily in the gauge 
where $A=\Ab =0$. Extensions to different gauges, e.g. the off-resonant 
case which requires a different gauge fixing as in Eq. (\ref{gfix}), 
can be made by the vector gauge transform in Eq. (\ref{gaugetr}). 
One peculiar property of the action in Eq. (\ref{action2}) is its asymmetry 
under the change of parity through $z \leftrightarrow \zb $. This is 
because the Wess-Zumino-Witten action in Eq. (\ref{wzw}) is a sum of the parity 
even kinetic term and the parity odd Wess-Zumino term thereby breaking 
parity invariance. In the optics context, broken parity is due to the 
slowly varying enveloping approximation which breaks the apparent parity 
invariance of the Maxwell-Bloch equation. Nevertheless, the action 
in Eq. (\ref{action2}) is invariant under the chiral transform
\be
z \leftrightarrow \zb \  
, \ g \leftrightarrow g^{-1} ~ (\mbox{ or } \h \leftrightarrow -\h 
\ , \ \vf \leftrightarrow -\vf )      .
\label{chiral}
\ee
This may be compared with the $CP$ invariance in  particle physics. 
Thus, parity invariance is in fact not lost but 
appears in a different guise, namely the chiral invariance. This 
chiral symmetry relates two distinct solutions, or it generates a new 
solution from a known one. For example, under the chiral transform in Eq. 
(\ref{chiral}), the 1-soliton solution in Eq. (\ref{solisol}) in the resonant 
case $(\xi = 0)$ becomes again a soliton but with the replacement of 
constants $a, b$ by  
\be
a \rightarrow -{a \over a^{2} + b^{2}} \ , \ 
b \rightarrow { b \over a^{2} + b^{2}}.
\ee
This implies the change of pulse shape and the change of pulse velocity 
by $v \rightarrow c - v$ (see Fig. 3). The current and the charge also change by
\be
J \rightarrow -\bar{J} \ , \ \bar{J} \rightarrow - J \ , \ Q 
\rightarrow -Q .
\ee
It is remarkable that the velocity changes from $v$ to $c-v$ unlike the 
usual parity change $v \rightarrow -v$. 
\vglue .2in

\begin{figure}
\vglue -.9in   
\leftline{\epsfysize 1.9 truein \epsfbox {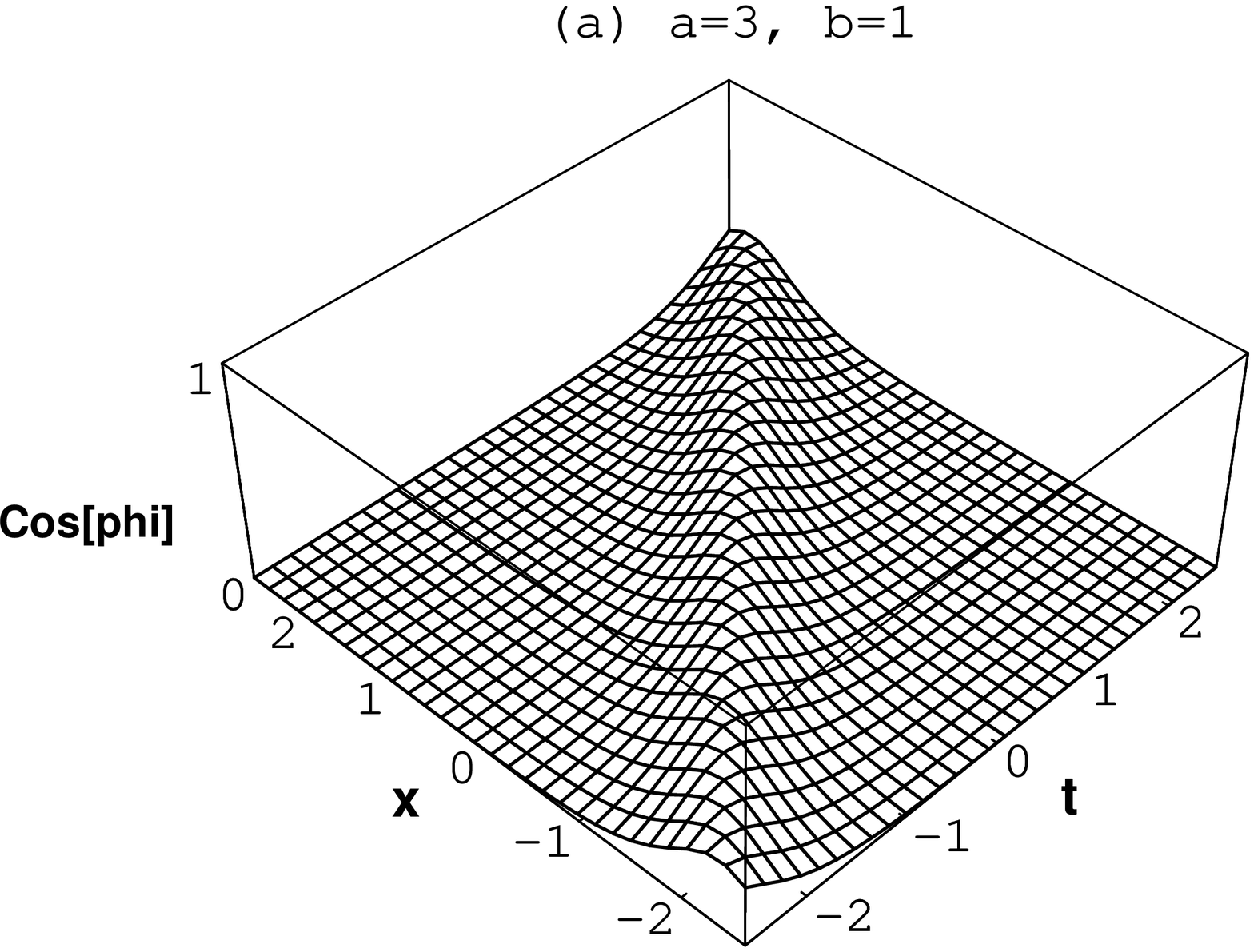}}
\vglue -1.9in
\rightline{\epsfysize 1.9 truein \epsfbox {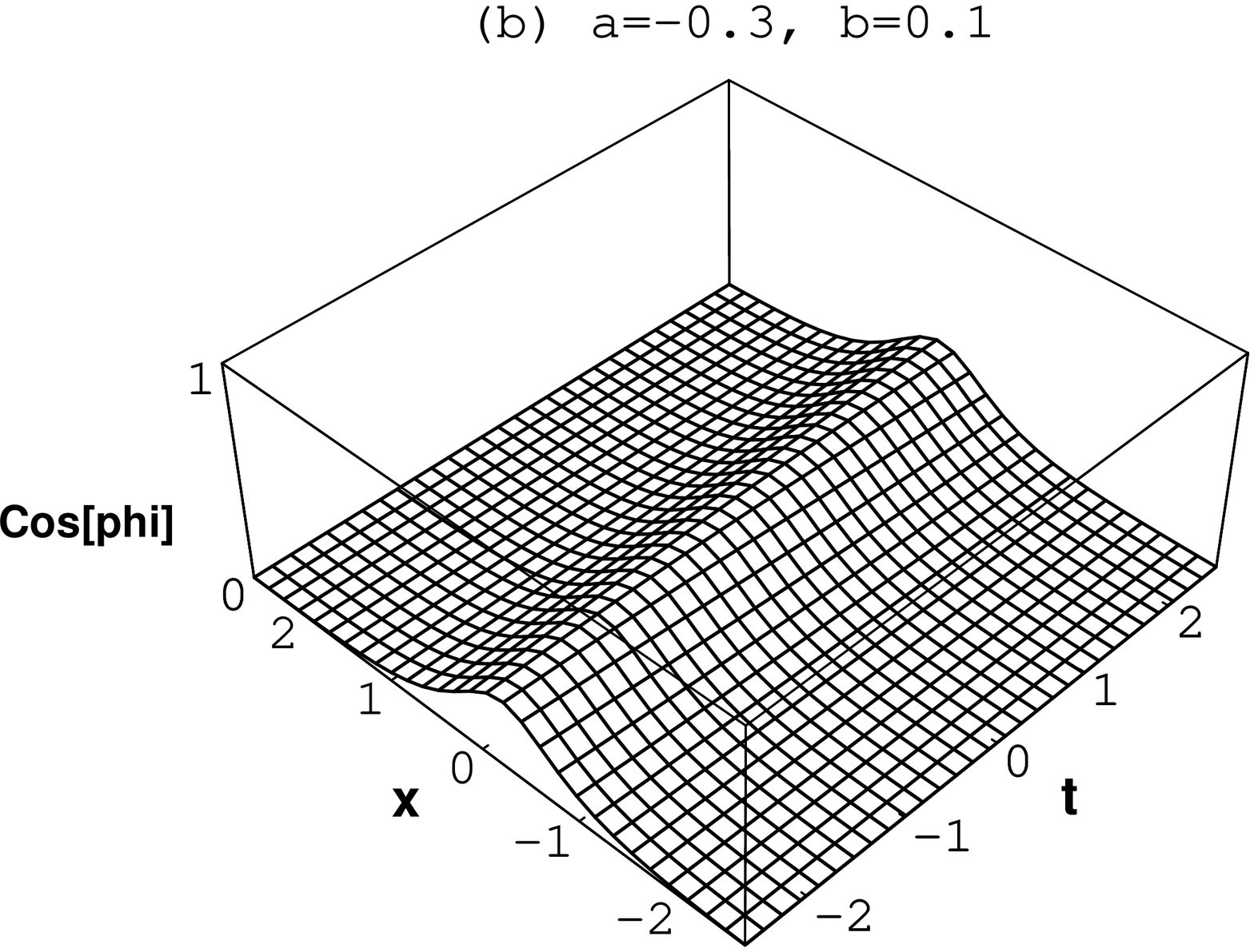}}
\caption{$\cos \vf$ for (a) one-soliton and (b) its chirally transformed soliton. }
\end{figure}

The other type of discrete symmetry of the action in Eq. (\ref{action2}) is
the dual symmetry: \cite{shin1}
\be
\b \leftrightarrow -\b \ , \ 
g \leftrightarrow  i\s g  
\ee
where $\s$ is a constant matrix with a property, $ \s T + T\s   =0$.
For example, $\s = \s_1$ of Pauli matrices in the $SU(2)/U(1)$ case.
This rather unconventional symmetry, also the name, stems from the 
ubiquitous nature of the action in Eq. (\ref{action2}), i.e. it also arises as a 
large level limit of parafermions in statistical physics and the above 
transform is an interchange between the spin and the dual spin 
variables \cite{park}. In general, the change of the sign of $\b $ 
makes the potential upside down so that the degenerate vacua becomes 
maxima of the potential and vice versa. Therefore, the dual transformed 
solutions are no longer stable solutions. This allows us to find a 
localized solution which approches to the maximum of the potential 
asymptotically (so called a ``dark" soliton). In practice, the dark 
soliton for positive $\b $ can be obtained by replacing $\b 
\rightarrow -\b \ , \ \zb \rightarrow -\zb $ in the ``bright" soliton 
of the negative $\b $ case. For example, we obtain the 
dark 1-soliton for the $SU(2)/U(1)$ case as follows:
\ben
\cos{\vf }e^{2i\h } &=& - {b \over \sqrt{(a-\x)^{2} + b^{2}}}
\mbox{tanh} (2bz + 2bC\zb ) - i{a-\xi \over \sqrt{(a-\xi )^{2} 
+ b^{2}}} \nonumber \\       
\q &=& -2(a- \x) (z- C\zb ) -2\xi z .
\label{darksol}
\een
Fig. 4 shows profiles of a dark soliton. Note that field intensity $|E|$ 
is the same as that of the bright soliton in Fig. 2. However, the 
population inversion $D$ for the dark soliton becomes inverted 
compared to that of the bright soliton. 
\begin{figure}
\vglue -.9in  
\leftline{\epsfysize 1.9 truein \epsfbox {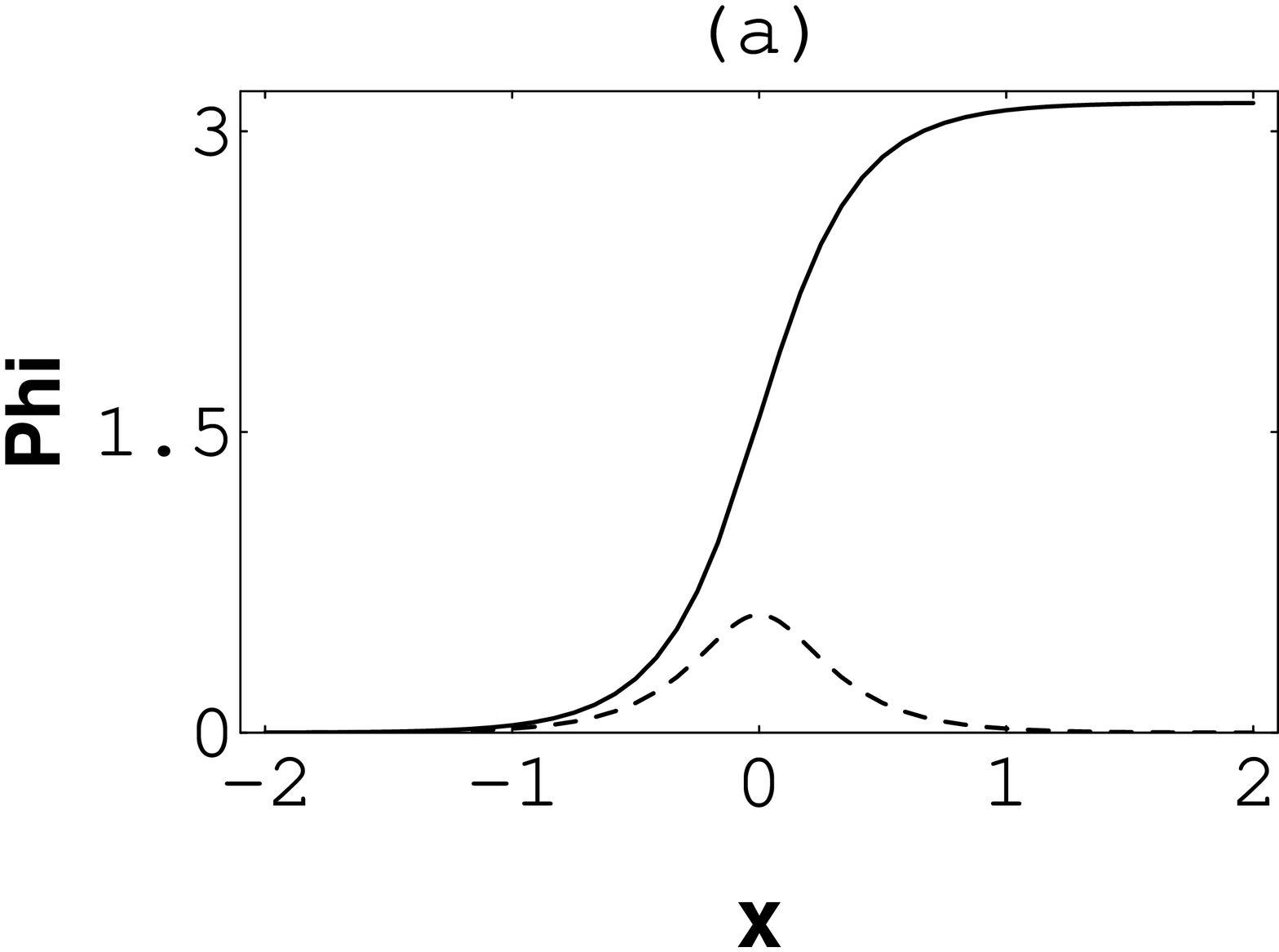}}
\vglue -1.9in
\rightline{\epsfysize 1.9 truein \epsfbox {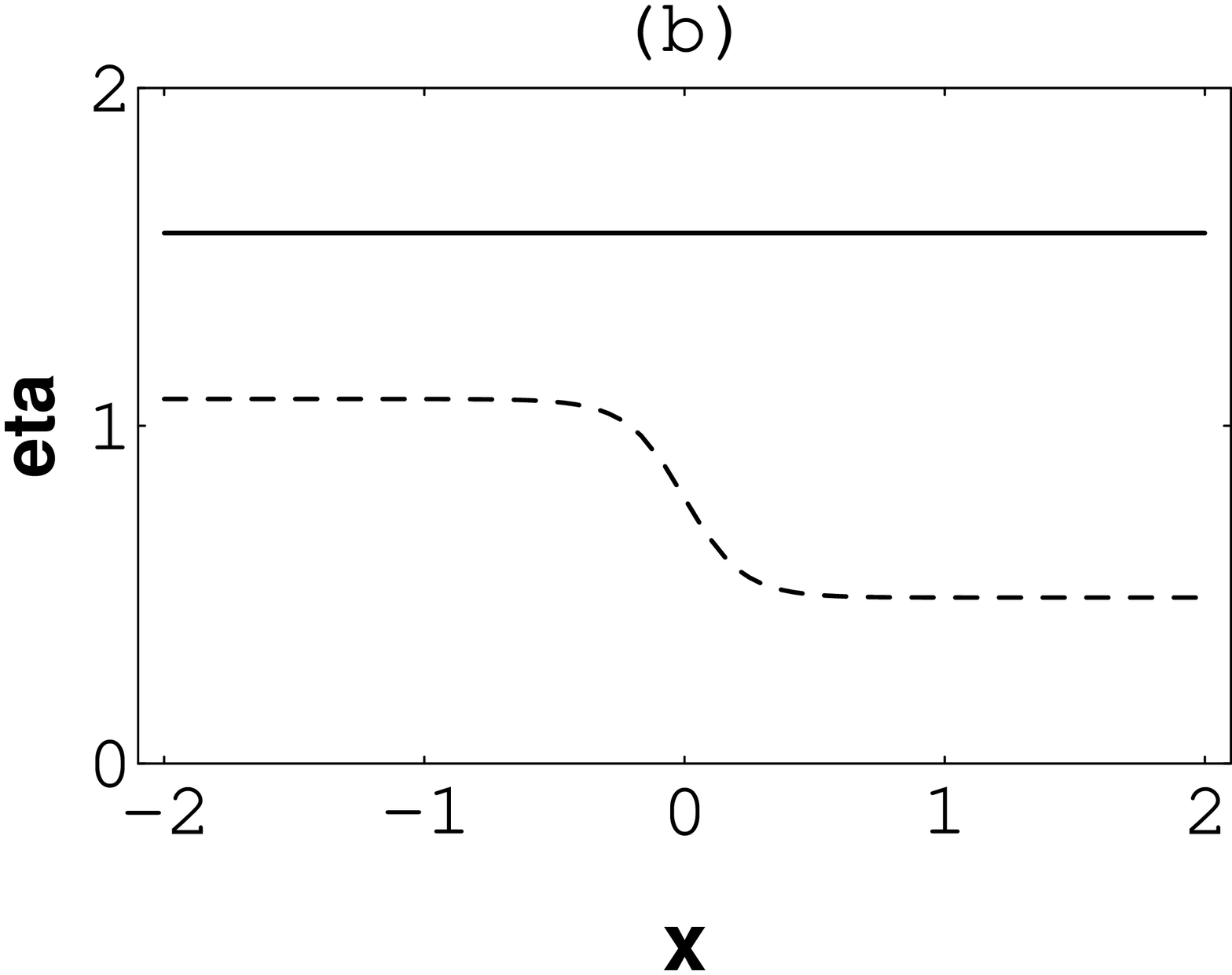}}
\vglue .3in
\leftline{\epsfysize 1.9 truein \epsfbox {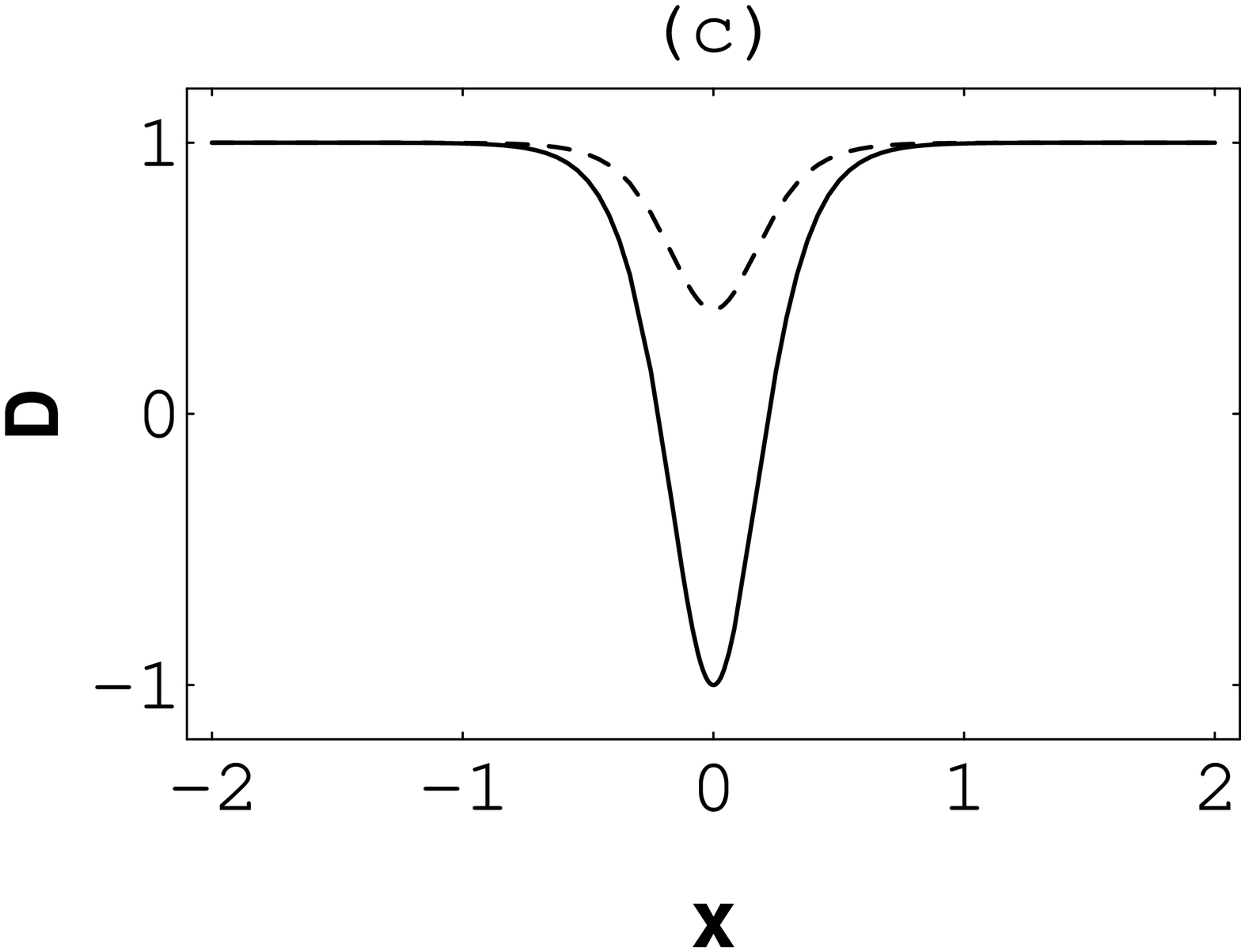}}
\vglue -1.9in
\rightline{\epsfysize 1.9 truein \epsfbox {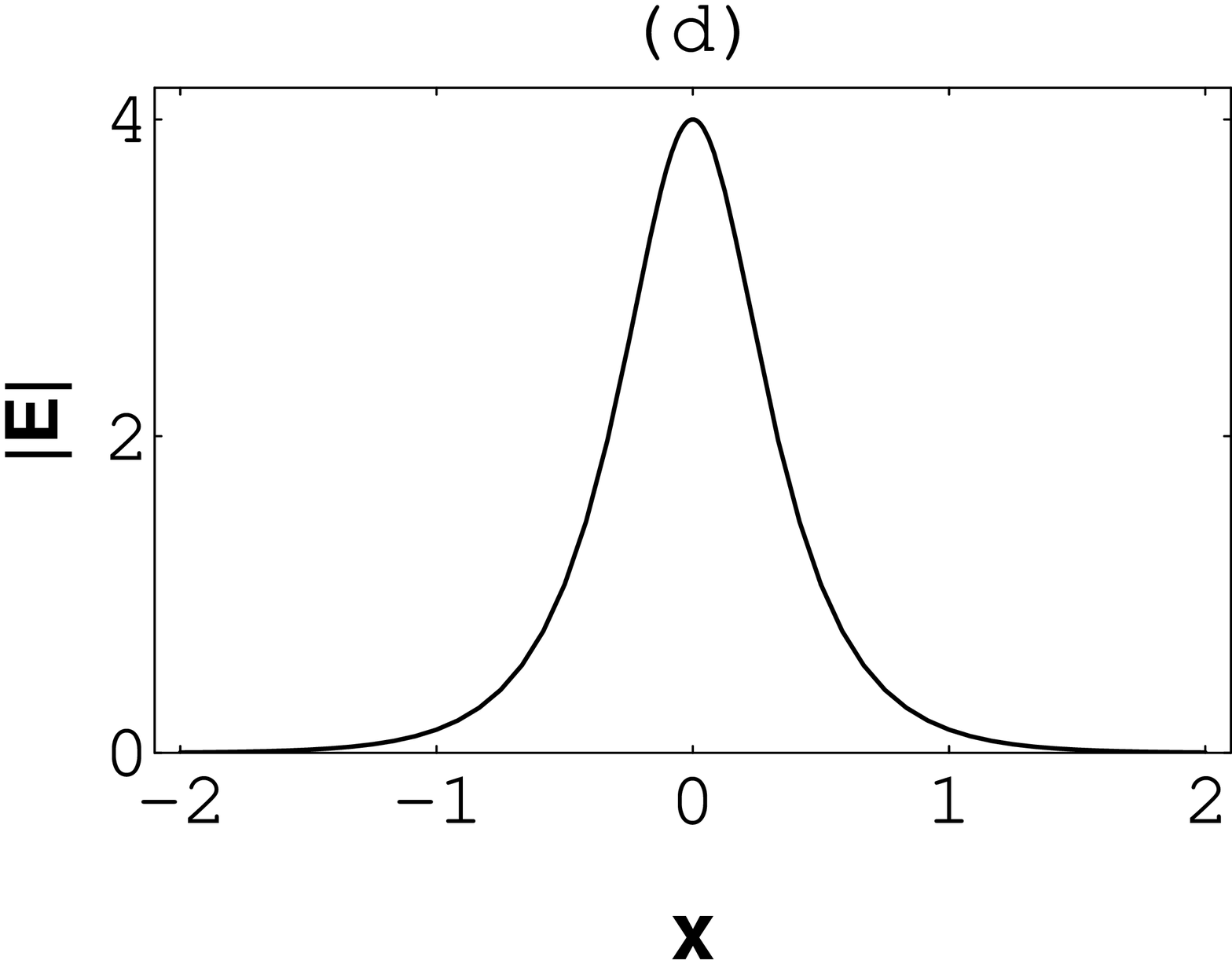}}                                            
\caption{Plots of $\f$,  $\q$ ,  $D$ and $|E|$ for dark one-soliton with 
$b=2 , ~ \x$=0 . 
The plot is w.r.t. $x= 2bz + 2bC_{1} \zb $. Real ($a=0$) and dashed ($a=3$) 
lines represent the topological and the nontopological solitons.
}
\end{figure}

\section{Discussion}
\setcounter{equation}{0}
In this paper, we have introduced a potential concept to optical systems 
described by the Maxwell-Bloch equation. In terms of the matrix potential, 
a field theory action for the Maxwell equation was established where the 
Bloch eqaution became a mere identity. Various identifications of 
multi-level systems have been made in associated with specific symmetric
spaces $G/H$ and the resulting group theoretic properties have been used in 
constructing conserved integrals. The field theory action uncovered several 
new features of the Maxwell-Bloch system; gauge symmetry, topological and 
nontopological charges, self-detuning, modified area theorem etc..
In doing so, the introduction of a matrix potential variable $g$ was an 
essential step. One immediate question is about the generality of such a 
potential variable in the description of nonlinear optics problems. 
Throughout the paper, we have confined ourselves only to the integrable 
Maxwell-Bloch equations which admit the inverse scattering method. Also, 
we have concentrated only on the classical aspect of the field theory  
which gives a semi-classical description of light-matter interaction.
In general, the Maxwell-Bloch equation is not integrable. Even the 
integrable cases require specific physical settings. For example, in 
the three-level system, integrability requires equal oscillator strengths.
However, we emphasize that integrability is not a necessary condition for 
the matrix potential formalism. Even for nonintegrable cases, one could 
still solve the Bloch equation in terms of a matrix potential $g$, and express 
the Maxwell equation in terms of $g$ \cite{ps3}. One example is 
the $2 \rightarrow 2 $ transition in the degenerate two-level system which is 
described by the double sine-Gordon equation when certain restrictions are 
made. A more general matrix potential treatment of nonintegrable cases will 
be considered elsewhere.      
\\

On the other hand, the group theoretic approach in terms of $g$ is not 
restricted to the Maxwell-Bloch systems only. The nonlinear Schr\"{o}dinger 
equation, which is the governing equation for optical soliton communication 
systems, can be generalized according to each Hermitian symmetric 
spaces \cite{fordy}. In fact, both the Maxwell-Bloch and the nonlinear 
Schr\"{o}dinger equations share the same Hamiltonian structure and they can 
be combined together. This case and its physical applications will be 
considered in a separate paper. Finally, we point out that our field theory 
formulation provides a vantage 
point to the quantum Maxwell-Bloch system as well as the quantum optics 
itself. A direct quantization of the Maxwell-Bloch equation using the quantum 
inverse scattering has been made by Rupasov and a localized multiparticle 
state has been found and compared with a quantum soliton \cite{rupasov}. 
Our field theory formulation suggests an alternative, yet more systematic 
way of quantization through the use of quantum field theory. 
The appearance of specific coset structures and their Hermitian properties 
suggests that a systematic quantization based on group theory is possible. 
Once again, this is not restricted to integrable cases and extensions to 
other quantum optical systems can be made. This work is in progress and 
will be reported elsewhere.

\vglue .3in 
{\bf ACKNOWLEDGEMENT}
\vglue .2in
This work was supported in part by Korea Science and Engineering 
Foundation (KOSEF) 971-0201-004-2, and by the program of Basic Science 
Research, Ministry of Education BSRI-97-2442, and by KOSEF through 
CTP/SNU.
\vglue .2in
\vglue .3in
{\bf APPENDIX A: Inverse scattering method and the matrix potential}   
\setcounter{equation}{0}
\renewcommand{\theequation}{A.\arabic{equation}}
\vglue .2in
In the following, we show that the matrix potential is intimately related 
to the dressing (inverse scattering) method and explain how to obtain 
exact solutions. The dressing method is a systematic way to obtain nontrivial 
solutions from a trivial one. In our case, we take a trivial solution by 
\be 
g = 1 \mbox{ and }
 \Psi = \Psi^{0} \equiv
 \exp[ - (A -\xi T + \lt T )z -  \left< { \bar{T}  \over \tilde{\l } - 
 \xi^{'} } \right>\bar{z} ] .
\ee
Let $\Gamma $ be a closed contour or a contour extending to infinity 
on the complex plane of the parameter $\lt $ and $G(\lt )$ be a matrix 
function on $\Gamma $.  Consider the Riemann problem of 
$\Psi^0 G(\lt )(\Psi^0)^{-1}$ on $\Gamma $ which consists of the 
factorization 
\be
\Psi^0 G(\lt )(\Psi^0)^{-1} = 
(\Phi_{-})^{-1} \Phi_{+}
\label{factor}
\ee
where the matrix function $\Phi_{+}(z,\zb ,\lt )$ is analytic with $n$ 
simple poles $\m_{1}, ... ,\m_{n}$ inside $\Gamma $ and 
$\Phi_{-}(z,\zb ,\lt )$ analytic with $n$ simple zeros 
$\l_{1}, ... , \l_{n}$ outside $\Gamma$. We assume that none of these 
poles and zeros lie on the contour $\Gamma$ and the factorization is 
analytically continued to the region where $\l \ne \m_{i} ,
\l_{i} \ ; \ i = 1,...,n$.
We normalize $\Phi_{+}, \Phi_{-}$ by  $\Phi_{+}|_{{\tilde \l} = 
\infty }= \Phi_{-}|_{{\tilde \l} = \infty } =1$. 
Differentiating Eq. (\ref{factor}) with 
respect to $z$ and $\zb $, one can easily show that
\ben
\pp \Phi_{+} \Phi_+^{-1} -  \Phi_{+} (A-\xi T + \lt T ) 
\Phi_{+}^{-1} &=& \pp \Phi_{-} \Phi_{-}^{-1} - \Phi_{-} (A - \xi T + 
\lt T ) \Phi_{-}^{-1}
\nonumber \\
\pb \Phi_{+} \Phi_+^{-1} - \left< { \Phi_{+} \Tb \Phi_{+}^{-1} \over 
\lt - \xi^{'} }  \right>  &=& \pb \Phi_{-} \Phi_{-}^{-1} - 
\left< { \Phi_{-} \Tb \Phi_{-}^{-1} \over \lt - \xi^{'} } \right>  .
\een
Since $\Phi_{+} (\Phi_{-})$ is analytic inside(outside) $\Gamma$, 
we find that the matrix functions $\bar{U} $ and $ \bar{V} $, defined by
\ben
\bar{U} &\equiv &  -\pp \Phi \Phi^{-1} +  \Phi (A - \xi T + \lt  T )
\Phi ^{-1} - \lt T
 \nonumber \\
\bar{V}  &\equiv &  -( \lt - \xi ) \pb \Phi \Phi^{-1} +  \Phi \Tb 
\Phi^{-1}
\een
where $\Phi = \Phi_{+}$ or $\Phi_{-}$ depending on the region, become
independent of $\lt $. Then, $\Psi \equiv \Phi \Psi^{0}$ satisfies the
linear equation;
\be
(\pp + \bar{U} + \lt T )\Psi = 0 \ , \ (\pb + \left< { \bar{V}  
\over \tilde{\l } - \xi^{'} }  \right> )\Psi = 0 \ .
\ee
Since $\bar{U}, \bar{V}$ are independent of $\lt $, we may fix $\lt $ by 
taking $\lt = \xi $. Define $g$ by $ g \equiv H\Phi^{-1} |_{\lt = \xi } $ 
where $H$ is an arbitrary constant matrix which commutes with 
$T, \bar T$ and $A$. Then, $\bar{U}$ and $ \bar{V}$ become
\ben
\bar{U} &=& g^{-1}\pp g +  g^{-1}Ag - \xi T  \\
\bar{V} &=&  g^{-1}\bar{T}g .
\een
If we further require the constraint condition (\ref{inhocon}) on 
$\Phi^{-1} |_{\lt = \xi } $ such that
\be
 (-\pp \Phi \Phi^{-1} +  \Phi A \Phi ^{-1})_{\bf h} -A =0,
\ee
we obtain a nontrivial solution $g$ and $\Psi $ from a trivial one. 
The nontrivial solution in general describes n-solitons coupled with 
radiation mode. If $G(\lt ) = 1$ in Eq. (\ref{factor}), we obtain exact 
n-soliton solutions. This formal procedure may be carried out 
explicitly for each cases of SIT in Sec. 3 and a closed form of 
n-soliton solutions can be obtained. 
\vglue .3in
{\bf APPENDIX B: Conserved local integrals} 
\setcounter{equation}{0}
\renewcommand{\theequation}{B. \arabic{equation}}
\vglue .2in
We first review some basic facts about Hermitian symmetric 
space \cite{helgason} which are relevant for the construction of 
conserved integrals. A symmetric space $G/H$ is a coset space with the 
Lie algebra commutation relations among generators of associated Lie 
algebras such that
\be
[ \bf{h} \ , \ \bf{h} ] \subset \bf{h} \ ,\ [ \bf{h} \ , \ \bf{m} ] 
\subset \bf{m} \ , \ 
[\bf{m} \ , \ \bf{m} ] \subset \bf{h} \ ,
\ee
where $\bf{g}$  and $\bf{h}$ are Lie algebras of $G$ and $H$ and 
$\bf{m}$ is the vector space complement of $\bf{h}$ in $\bf{g}$, i.e. 
\be
\bf{g} = \bf{h} \oplus \bf{m}.
\label{decom}
\ee
Hermitian symmetric space is a symmetric space equipped with a 
complex structure. In general, such a complex structure 
is given by the adjoint action of $T_{0}$ on ${\bf m}$ up to a scaling, 
where $T_{0}$ is an element belonging to the Cartan subalgebra of 
${\bf g}$ whose stability subgroup is $H$. In our case, $T_{0}$ is 
precisely the $T$-matrix given in Sec. 3. Namely, with a suitable 
normalization of $T$, we have
\be
T \in {\bf h} \ \  , \ \ [T\ , \ {\bf h}]=0 \ \ \mbox{ and } 
\ \ [T\ ,\ [T\ ,\ a]]= - a \ \mbox{ for any } a \in {\bf m}.
\ee
We decompose an algebra element $\j \in {\bf g} $ according to 
Eq. (\ref{decom}),
\be
\j = \j_{h} + \j_{m} .
\ee
Such a decomposition could be extended to a representation $\Psi$ 
of $ G = SU(n)$ if we substitute the commutator by a direct matrix 
multiplication and add an identity element $h_{0} = I $ to the 
subalgebra ${\bf h}$, i.e.
\be
\J = \J_{h} + \J_{m} , ~~~ \J_{h} \J_{h} \subset \J_{h} , ~~ 
\J_{h}\J_{m} \subset \J_{m} , ~~ \J_{m} \J_{m} \subset \J_{h} .
\ee
In other words, any unitary $n \times n$ matrix can be expressed 
as a linear combination of $SU(n)$ generators and the identity element 
$h_0$ such that 
\be
\J = \J_{h} + \J_{m} = \sum_{a=0}^{\mbox{dim } \bf h} \J_{h}^{a} h_{a} + 
\sum_{b=1}^{\mbox{dim } \bf m} \J_{m}^{b} m_{b} .
\ee
In order to solve the linear equation (\ref{inholin}) recursively, we 
expand $\J$ in terms of a power series in $ {\l}$,
\be
\J \exp (-{ {\l}} T z) =\sum _{i=0} ^{\infty} {1 \over { 
{ \l}}^{i} } \Phi_{i}, 
\ee
where 
\be
\F_i = \sum_{a=0}^{\mbox{dim } \bf h} \F_{hi} ^{a} h_{a} + 
\sum_{b=1}^{\mbox{dim } \bf m} \F_{mi}^{b} m_{b} \equiv \cC _{i} + 
\cD _{i} .
\ee 
With the notation in Eq. (\ref{ghabb}), the linear equation is given by
\be
(\pp + \E ) \F_{i} - [T\ ,\ \F_{i+1}]=0
\label{itcon1}
\ee
and 
\be
\pb \F_{i} + \sum_{l=0}^{i-1} (D_{i-l-1} + P_{i-l-1}) \F_l =0.
\label{itcon2}
\ee
Then, the $\bf{m}$-component of Eq. (\ref{itcon1}) is
\be
\pp \cD _{i-1} + \E \cC _{i-1} -[T\ ,\ \cD _{i}]=0 ,
\ee
which can be solved for $ \cD _{i}$ by applying the adjoint action 
of $T$, 
\be
\cD _{i} = - [T\ ,\ \pp \cD _{i-1}]-[T\ ,\ \E ]\cC _{i-1}.
\ee
$\cC _{i}$ can be solved similarly from the $\bf{h}$-component 
of Eqs. (\ref{itcon1}) and (\ref{itcon2}) such that
\be
\cC _{i} = - \int \E \cD _{i} dz 
- \sum_{l=0}^{i-1} \int (D_{i-l-1} \cC _l + P_{i-l-1} \cD _l ) d \bar z.
\ee  
Finally, the conserved current follows from the consistency condition 
$\pp \pb \cC _{i} = \pb \pp \cC _{i}$. 

With the repetitive use of the properties of the Hermitian symmetric 
space, it can be easily checked that these conservation laws are indeed 
consistent with the equations of motion (\ref{inhozero}) and (\ref{ident}), 
which in the present convention take a particularly simple form:
\ben
\pb \E + [T ~ , ~ P_0] &=& 0 \nonumber \\
\pp D_{i} + [\E ~,~ P_{i}] &=& 0 \nonumber \\
\pp P_{i} + [\E ~, ~ D_{i} ]-[T\ ,\ P_{i+1}] &=& 0.
\een
In general, the conserved current contains nonlocal terms. 
These nonlocal terms may be dropped out by taking the $T$-component of 
the currents. For instance, the $T$-component of the ``spin-2" current 
conservation is
\be
\pb {\rm Tr} (T \E \pp \E ) = \pp {\rm Tr} (T P_0[T\ ,\ \E ]-T D_1)
\ee
which obviously does not contain nonlocal terms.

\end{document}